*Data and text mining*

# ModuLand plug-in for Cytoscape: determination of hierarchical layers of overlapping network modules and community centrality

Máté Szalay-Bekő[1,‡] and Robin Palotai[1,‡], Balázs Szappanos[2], István A. Kovács[1,3], Balázs Papp[2] and Péter Csermely[1,*]

[1]Department of Medical Chemistry, Semmelweis University, Budapest, Hungary

[2]Institute of Biochemistry, Biological Research Centre, Hungarian Academy of Sciences, Szeged, Hungary

[3]Department of Physics, Loránd Eötvös University, Budapest, Hungary and Institute for Solid State Physics and Optics, Wigner Research Centre, Hungarian Academy of Sciences, Budapest, Hungary

*To whom correspondence should be addressed. E-mail: csermely.peter@med.semmelweis-univ.hu  ‡ equal contributions

**ABSTRACT**

**Summary:** The ModuLand plug-in provides Cytoscape users an algorithm for determining extensively overlapping network modules. Moreover, it identifies several hierarchical layers of modules, where meta-nodes of the higher hierarchical layer represent modules of the lower layer. The tool assigns module cores, which predict the function of the whole module, and determines key nodes bridging two or multiple modules. The plug-in has a detailed JAVA-based graphical interface with various colouring options. The ModuLand tool can run on Windows, Linux, or Mac OS. We demonstrate its use on protein structure and metabolic networks.

**Availability:** The plug-in and its user guide can be downloaded freely from: http://www.linkgroup.hu/modules.php.

**Contact:** csermely.peter@med.semmelweis-univ.hu

**Supplementary information:** Supplementary information is available at *Bioinformatics* online.

## 1 INTRODUCTION

Nodes of biological networks often belong to multiple network communities. Recently, a number of methods were published to determine tightly or extensively overlapping network modules (Adamcsek *et al.*, 2006; Ahn *et al.*, 2010; Fortunato, 2010; Kovacs *et al.*, 2010; Mihalik and Csermely, 2011; Palla *et al.*, 2005). Our ModuLand framework (Kovacs *et al.*, 2010) introduced community landscapes. The *x-y* plane of a community landscape is a conventional 2D visualization of the network, while the *z* axis represents community centrality. Community centrality of a given edge (or node) was defined as the sum of local influence zones of all network edges (or nodes) including the given edge (or node; Suppl. Figure 1). Thus community centrality represents an integrated measure of the whole network's influence to one of its edges or nodes. Hills of the community landscape correspond to network modules (Suppl. Figure 1) yielding extensive overlaps. This concept led to the development of the ModuLand family of network modularization methods (Kovacs *et al.*, 2010).

The widely used Cytoscape program (Shannon *et al.*, 2003) has several very useful clustering plug-ins (Bader and Houge, 2003; Morris *et al.*, 2011; Rhrissorrakrai and Gunsalus, 2011; Rivera *et al.*, 2010; Su *et al.*, 2010). However, these methods do not focus on extensive modular overlaps, and do not build a modular hierarchy, where meta-nodes of the higher level represent modules of the lower level. Moreover, existing plug-ins do not provide measures identifying the centre of the module, as well as key nodes bridging two or multiple modules (see Suppl. Table 9, and Suppl. Discussion). Here, we introduce the Cytoscape plug-in of the most widely applicable version of the ModuLand method family (Kovacs *et al.*, 2010). We demonstrate its ability to determine biologically relevant, extensively overlapping network modules, hierarchical layers of modules, module cores and key inter-modular nodes using protein structure and metabolic networks.

## 2 SOFTWARE OVERVIEW

The ModuLand Cytoscape plug-in uses the LinkLand influence zone determination method and the ProportionalHill module assignment method of our formerly published ModuLand network module determination method family (Kovacs *et al.*, 2010). These two methods provide a good trade-off between the fast (but rather inaccurate), and accurate (but rather slow) other ModuLand methods.

The installation of the ModuLand plug-in follows Cytoscape procedures. This is much easier than the setup required for the earlier version (Kovacs *et al.*, 2010). The program can be distributed as a single .jar file. Moreover, the current implementation works on Linux, Windows and Mac OS, thereby extending the options of the former version.

The plug-in determines extensively overlapping modules using any undirected network type and weight description of Cytoscape. Moreover, the plug-in calculates a set of hierarchical modules. In the modular hierarchy, modules of the lower level become meta-nodes of the upper level, and modular overlaps of the lower level become weights of the meta-edges at the upper level. The plug-in creates, automatically re-loads and visualizes the higher and higher level hierarchies (with lower and lower number of meta-nodes and meta-edges, see Figure 1), until the whole network coalesces into a single meta-node.

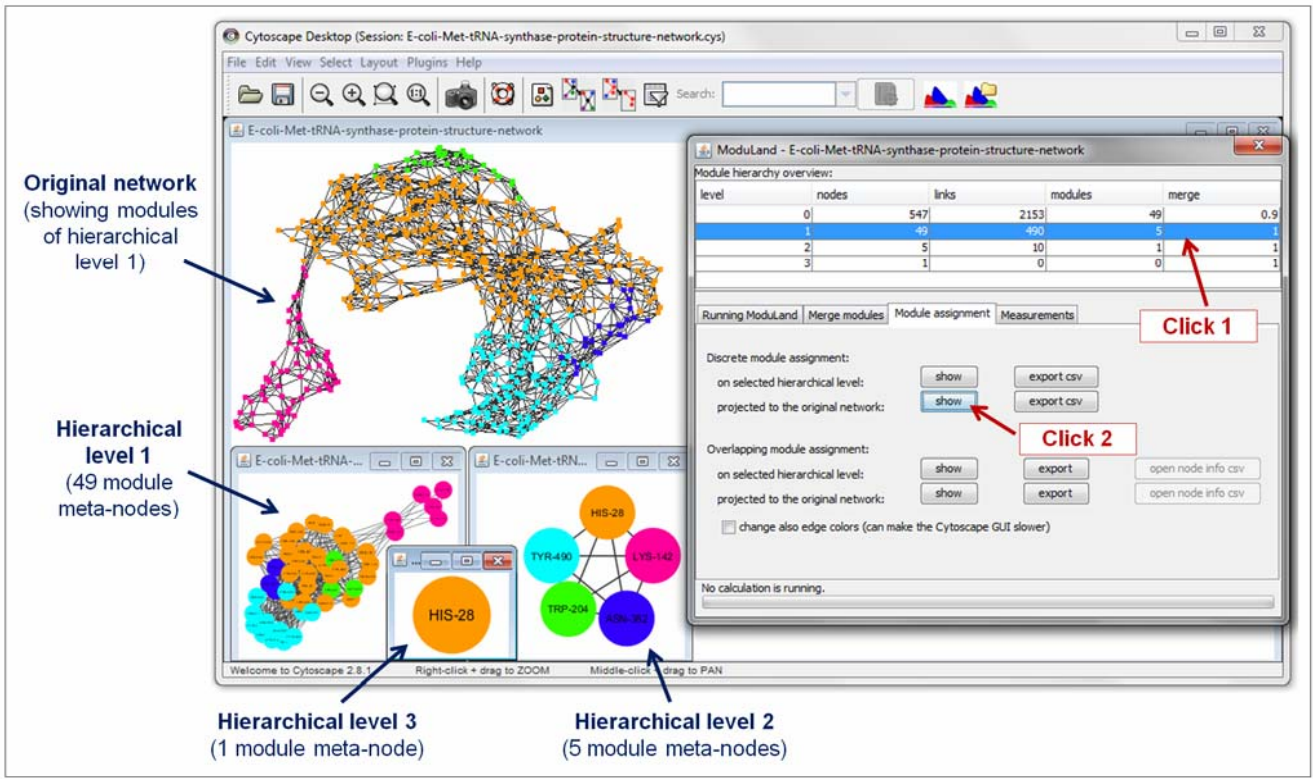

**Fig. 1.** The hierarchical modular structure determined by the ModuLand Cytoscape plug-in. The left side shows the protein structure network of *E. coli* Met-tRNA synthase and its 3 hierarchical levels as determined by the plug-in. Each meta-node of a higher hierarchical level represents a module of the level right below. All networks are coloured according to the 5 modules identified on hierarchical level 1. This colouring option can be performed by the 2 clicks of the plug-in main dialog box shown on the right.

The lower number of modules at higher hierarchical levels may be visualized either using the meta-nodes of the higher hierarchical level itself, or projecting this higher level modular structure to the nodes of the original network. On any level of module hierarchy, nodes or meta-nodes can be visualized assigning them the colour of the module they mostly belong to. This shows a non-overlapping assignment of nodes to modules. Nodes can also be marked by blending the colours of the modules proportional to the overlapping module assignment of the given node. Edges may be optionally visualized as a mix of the colour of their two nodes. The plug-in sets meta-node labels on the higher hierarchical level based on the modules of one level below in the hierarchy. The meta-node on the higher hierarchical level representing the module has the name of the node having the highest modular assignment value for the corresponding module at one level below in the hierarchy.

Node colours can also visualize several node (or meta-node) measures including weighted degree, betweenness centrality, community centrality, overlap and bridgeness (see Suppl. Figure 2).

The plug-in has an option to merge highly similar module pairs containing roughly the same nodes or meta-nodes with the same intensity. For merging of modules the plug-in offers a correlation histogram, and allows the user to select an appropriate correlation threshold. The runtime complexity of the plug-in version remained $\sim O(n^3)$, as defined earlier (Kovacs et al., 2010). To enhance the performance of the plug-in for calculating the higher hierarchical layers further, we introduced a user-selected optimization. This results in the disappearance of meta-edges with very small weights at the higher hierarchical levels. These low intensity meta-edges are derived from the minor overlaps of distant modules of the lower level. This optimization allowed a speedup in running time by a factor of 7 for larger networks (Suppl. Table 10).

The plug-in is capable to generate overview reports for each hierarchical level. These reports list the number of the nodes (meta-nodes), edges (meta-edges) and modules, the effective number of modules (see Kovacs *et al.*, 2010) and the size of each module. The overview also contains the list of the 10 nodes of each module having maximal module assignment value to the respective module (called as the module core). Data related to the module assignment and the calculated measures of nodes (and meta-nodes of higher hierarchical levels) can be exported in a csv or txt format.

The plug-in contains a Help function, and a detailed step-by-step User Guide can also be downloaded from the plug-in webpage: www.linkgroup.hu/modules.php.

## 3 RESULTS AND CONCLUSION

ModuLand-derived communities of various yeast protein-protein interaction networks gave a functionally meaningful description of the yeast interactome (Kovacs et al., 2010). Function of module core proteins proved to be good indicator of the function of the whole module (Mihalik and Csermely, 2011). Here, we demonstrate the use of the ModuLand Cytoscape plug-in on the protein structure network of *E. coli* Met-tRNA synthase, since an elegant study (Ghosh and Vishveshwara, 2007) showed the existence of 4 alternative communication paths in this enzyme. The 5 major sub-domains of Met-tRNA synthase were well reflected by the 5 modules obtained at the second hierarchical level of the protein structure network (Figure 1; Suppl. Table 3). Key amino acids of the most frequently used communication path (Ghosh and Vishveshwara, 2007) either belonged to the module cores of the 3 modules involved in transmission of conformational changes, or were inter-modular nodes between these modules (see Suppl. Table 4). These observations were in agreement with earlier findings (Ghosh and Vishveshwara, 2008; Sethi *et al.*, 2009).

We further demonstrated the use of the ModuLand plug-in by comparing the modular structures of the metabolic networks of the free-living bacterium *E. coli* and the endosymbiont *B. aphidicola* (Pál *et al.*, 2006). *E. coli* metabolic module cores had a significant overlap (Fisher's exact test p=1.4 x $10^{-7}$; see Suppl. Information for more details) with the modules determined earlier by Guimera and Amaral (2005).

Both visual inspection (see Suppl. Figures 4 to 7) and numerical values (see Suppl. Table 7) suggested a more differentiated modular structure of the *E. coli* metabolic network than that of *B. aphidicola*. This finding is in agreement with earlier findings (Kreimer *et al.*, 2008; Mihalik and Csermely, 2011; Parter *et al.*, 2007; Samal *et al.*, 2011). The difference in modular structure was not likely to be caused by the difference in the size of the *E. coli* and *B. aphidicola* networks (see Suppl. Figures 8 and 9, and Suppl. Tables 7 and 8).

*E. coli* module cores corresponded to significantly less metabolic functions than those of *B. aphidicola* (0.53 *versus* 0.67 functions per module core reactions, respectively; bootstrap method p=0.0392). This difference remained even when we used an ensemble of 1000 randomly selected sub-networks of the *E. coli* metabolic network having the same number of nodes or edges as found in the smaller *B. aphidicola* network (see Suppl. Information for more details). Moreover, additional tests suggested that the large twin-modules forming the centre of the *B. aphidicola* network were not responsible for the differences observed in the number of metabolic functions (see Suppl. Information). These results indicated that modules of the metabolic network of an organism from a variable environment (*E. coli*) are more specialized than metabolic network modules of a symbiont having a constant environment (*B. aphidicola*). It is noteworthy that our result is in agreement with earlier findings using non-overlapping modularization (Parter et al., 2007), which is a further indication that the module cores of the plug-in capture well the biologically relevant function of modules.

In conclusion, the ModuLand Cytoscape plug-in provides a user-friendly and efficient method to identify and visualize a hierarchy of extensively overlapping modules, and determines key network positions (like module cores and bridges). As shown by several case studies, modules identified by the plug-in correspond to biologically meaningful groups, module cores help the identification of biological functions, and inter-modular nodes have a key role in a variety of biological networks.

## ACKNOWLEDGEMENTS

The authors thank the Editor and the anonymous Reviewers for their comments and suggestions, Amit Ghosh (Lawrence Berkeley National Laboratory, Berkeley CA, USA) and Saraswathi Vishveshwara (Indian Institute of Science, Bangalore, India) for the 3D coordinates of *E. coli* Met-tRNA-synthase, and members of the LINK-Group (www.linkgroup.hu) for their discussions and help. *Funding*: This work was supported by the EU [grant numbers FP6-518230, TÁMOP-4.2.2/B-10/1-2010-0013, FP7-264780], by the Hungarian Scientific Research Fund [grant numbers OTKA K-83314 and PD-75261], by the International Human Frontiers Science Program Organization (BP), by the Momentum Program of the Hungarian Academy of Sciences (BP) and by a residence at the Rockefeller Foundation Bellagio Center (PC). *Conflict of interest*: none declared.

**Supplementary Information for**

# ModuLand plug-in for Cytoscape: determination of hierarchical layers of overlapping network modules and community centrality[1]

Máté Szalay-Bekő[1,†] and Robin Palotai[1,†], Balázs Szappanos[2], István A. Kovács[1,3], Balázs Papp[2] and Peter Csermely[1,*]

[1]Department of Medical Chemistry, Semmelweis University, Budapest, Hungary

[2]Evolutionary Systems Biology Group, Biological Research Centre, Hungarian Academy of Sciences, Szeged, Hungary

[3]Department of Physics, Loránd Eötvös University, Budapest, Hungary and Research Institute for Solid State Physics and Optics, Budapest, Hungary

**See the supporting website for further information:**
http://www.linkgroup.hu/modules.php

---

[1] It is worth to note that the community centrality measure we use is not the same community centrality measure introduced by Mark Newman [2006]. The two community centralities are similar in the sense that they take into account the mesoscopic (modular) structure of the network to define the centrality value. However, the Newman-type community centrality is derived from eigenvector analysis, while ours represents the sum of local influence zones of all nodes or edges on each node or edge.

[†] These authors wish it to be known that, in their opinion, the first two authors should be regarded as joint First Authors.

# Contents

# Supplementary Figures

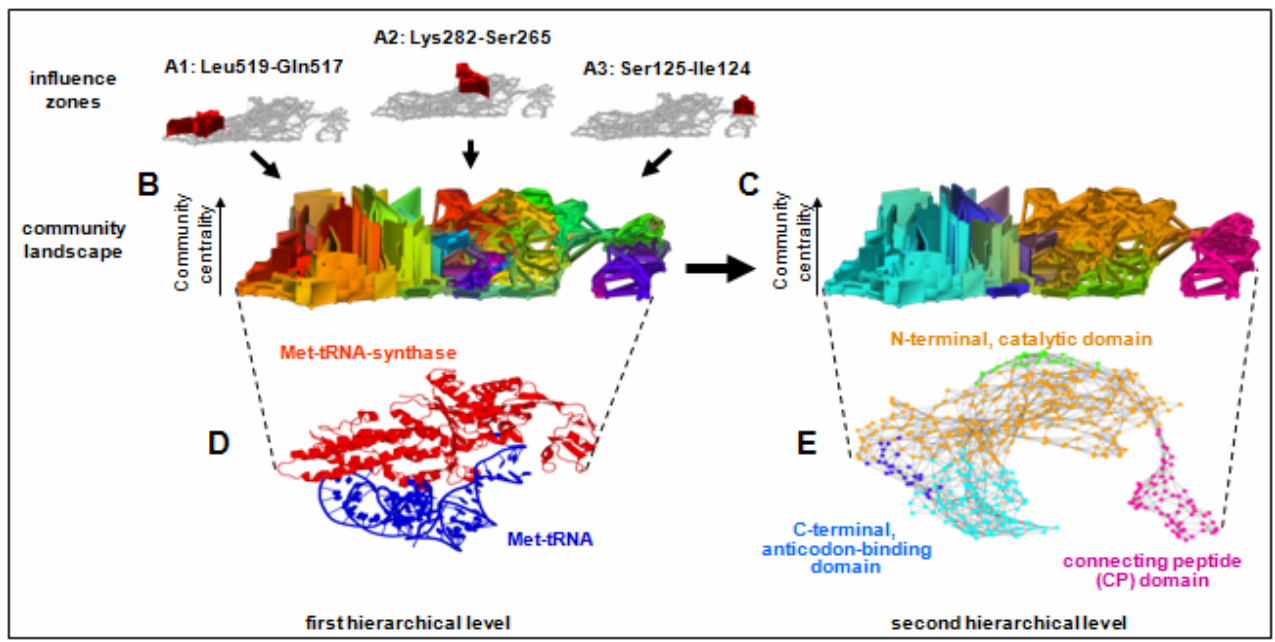

**Supplementary Figure 1. Hierarchical modules of Met-tRNA synthase protein structure network.** The figure illustrates the steps of modular analysis by the ModuLand Cytoscape plug-in on the example of the protein structure network of *E. coli* Met-tRNA synthase. The 3D positions of the amino acids and the protein structure network were constructed by [Ghosh and Vishveshwara, 2007] as described in Supplementary Methods. Panels A1, A2 and A3 show the influence zones of the edges between the three amino acid pairs indicated. Influence zones show the segment of the network affected by the given edge, and were calculated by the LinkLand algorithm as described previously [Kovacs *et al*., 2010]. Panel B shows the community landscape of the protein structure network. The height of the community landscape is the sum of all influence zones containing the given node or edge. This measure is called as community centrality. Hills of the community landscape correspond to the 49 overlapping modules of the first hierarchical level of the protein structure network shown by different colours. Modules were determined using a merge correlation threshold of 0.9. On Panel C the colours of the same community landscape represent the module structure of the second hierarchical level containing only 5 modules. Panel D shows the 3D structure of Met-tRNA (blue) and Met-tRNA synthase (red) aligned to the protein structure network. On the protein structure network of Panel E nodes and edges were coloured according to their assignments to the second hierarchical level. The three major domains of Met-tRNA-synthase (see Supplementary Table 1) are marked with labels coloured according to the colour of the most correlating module (see Supplementary Table 2B). The illustration on Panel D was made by the JMol program [Herráez, 2006]. Panel E represents a Cytoscape plug-in visualization using the organic layout option. Other Panels were created using the Blender program[2].

---

[2] Blender is the free open source 3D content creation suite, available for all major operating systems under the GNU General Public License. For more information see http://www.blender.org

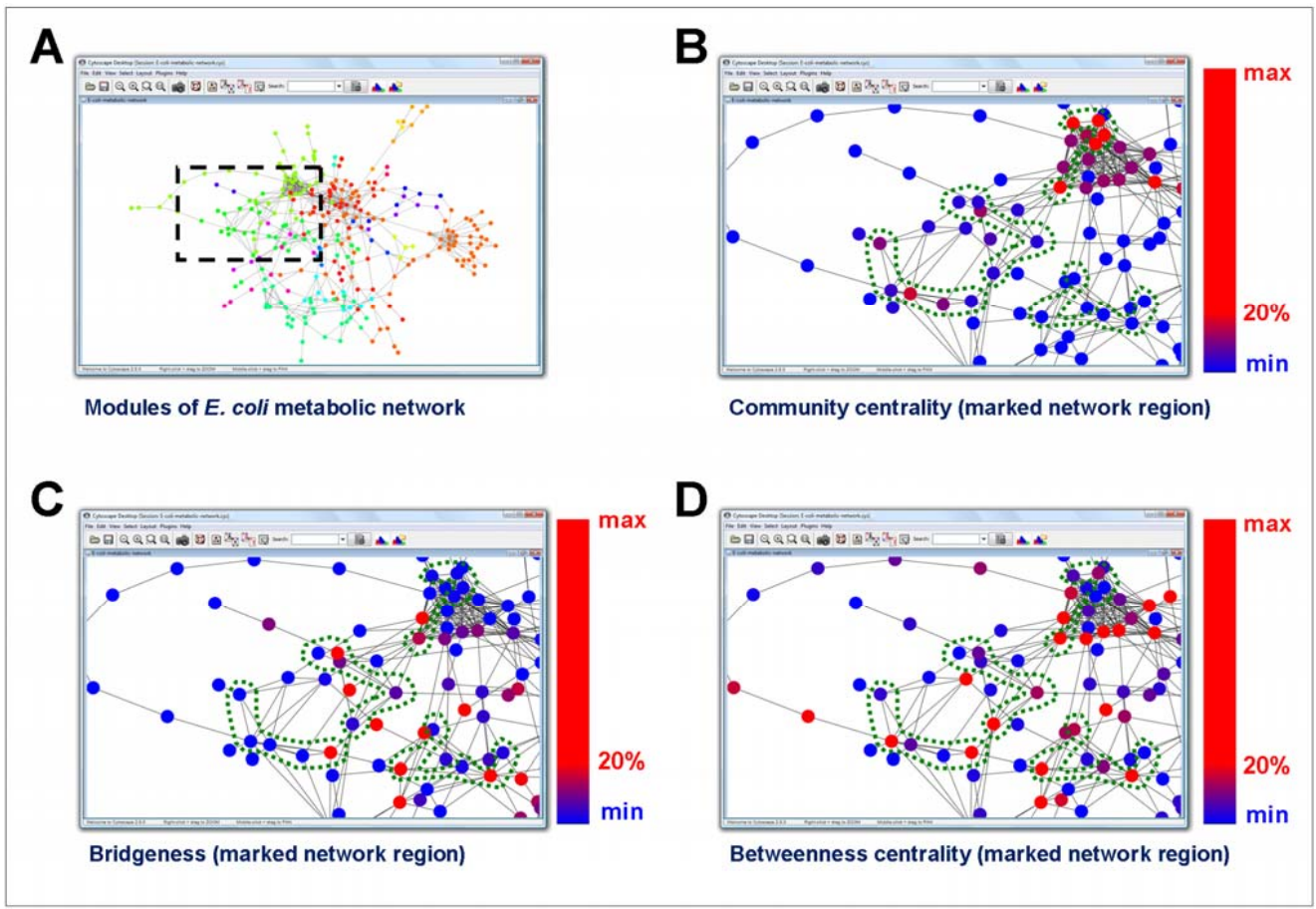

**Supplementary Figure 2. Illustration of a few centrality measures of the ModuLand Cytoscape plug-in.** The figure shows 4 screenshots of the E. coli metabolic network obtained by the plug-in. Panel A shows the 23 modules of the original network. Panels B, C and D show the community centrality, bridgeness and betweenness centrality measures of the smaller region of the network marked by dashed black square on Panel A. Measures were visualized by continuous colour mapping using blue-to-red colour scales on the VizMapper panel of Cytoscape from 0 to 20 percentages. Green dots highlight the top 5 core metabolites of the 4 major modules present in the marked network region. All networks were visualized using the Cytoscape Organic yLayout.

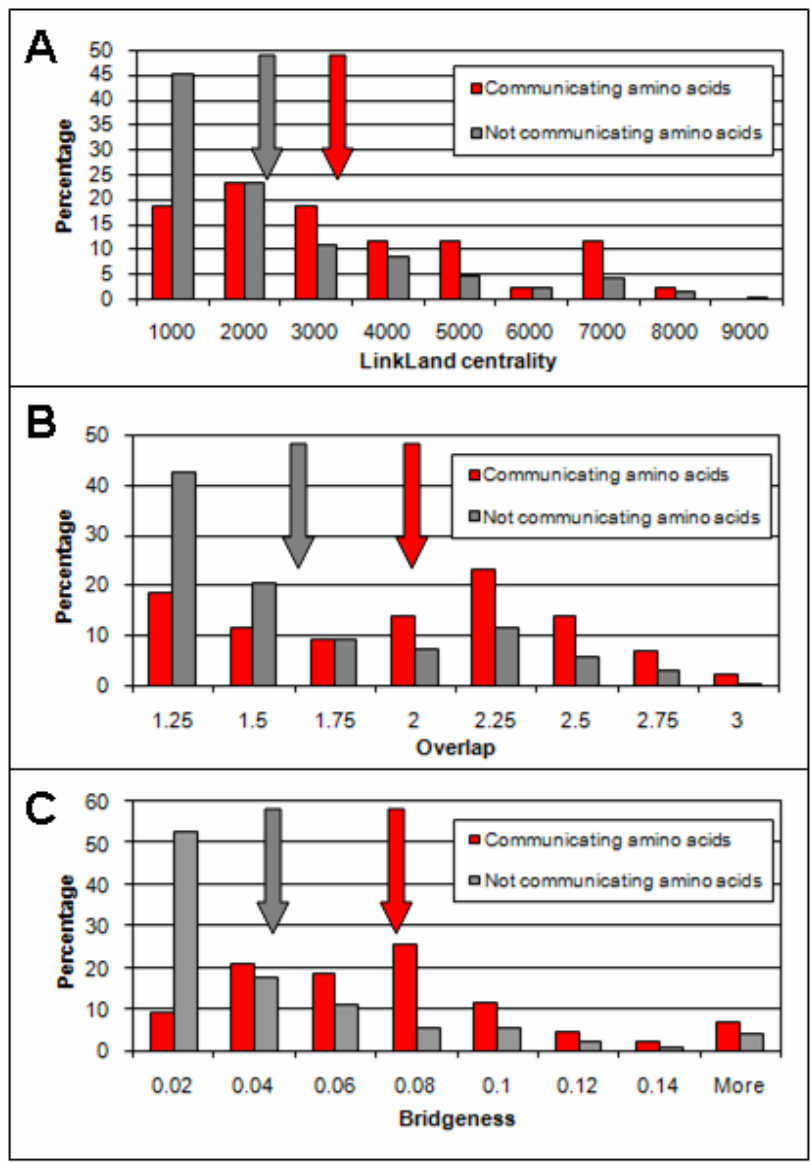

**Supplementary Figure 3. Community centrality, overlap and bridgeness of communicating amino acids of Met-tRNA synthase.** The protein structure network of *E. coli* Met-tRNA synthase was constructed as described in Supplementary Methods. Modules of the second hierarchical level were determined using the ModuLand Cytoscape plug-in with a merge correlation threshold of 0.9. Data of the 43 communicating amino acids participating in the transmission of conformational changes between the catalytic centre and the anticodon binding site of Met-tRNA synthase obtained from cross-correlations of molecular dynamics simulations [Ghosh and Vishveshwara, 2007], or data of the remaining 504 amino acids of the protein structure network are shown by red or grey bars, respectively. Average values were marked by the vertical arrows of the respective colours. Panels A through C show the community centrality, modular overlap and bridgeness values, respectively. The difference between the two datasets was verified using the Welch's two sample t-test (p-value < 0.0008 for all the three measurements). Community centrality values were determined using the LinkLand algorithm on the original network. The overlap values were calculated from module assignment values at the second hierarchical level. Bridgeness values represent the smaller of the two modular assignments of a node in two adjacent modules, summed up for every module pairs. This value is high, if the node belongs more equally to two adjacent modules in many cases, *i.e.* if it behaves as a bridge between a single pair, or between multiple pairs of modules. Such bridging positions correspond to saddles between the 'community-hills' of the 3D community landscape shown on Supplementary Figure 1B. Note that community centrality shows the influence of the rest of the network to the

given node, modular overlap reveals the simultaneous involvement of the node in multiple modules, while bridgeness characterizes an inter-modular position of the node between adjacent modules [Kovacs *et al.*, 2010].

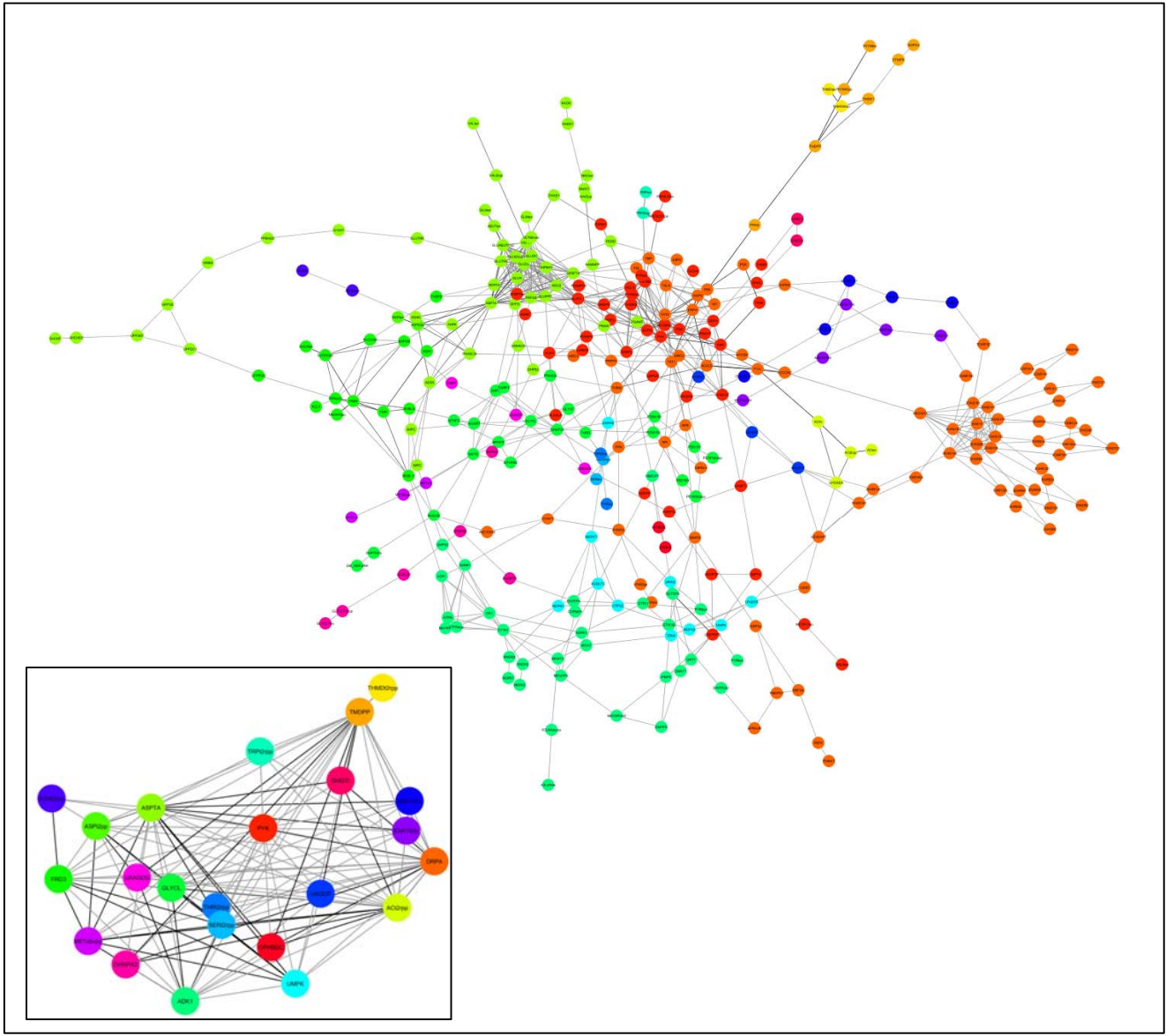

**Supplementary Figure 4. Hierarchical modules of *E. coli* metabolic network.** The construction of the network is described in Supplementary Methods. The inset shows the first level of module hierarchy of the network created by the ModuLand Cytoscape plug-in. Nodes in the inset represent modules on the original network. Images were created by the Cytoscape program [Shannon *et al*., 2003] in case of both networks, using the Organic yLayout. Edges were coloured in greyscale according to their weights. The colours of the nodes were set by the ModuLand Cytoscape plug-in. Nodes of the original network were coloured according to the colour of the module, where they mostly belong. In the first hierarchical level shown at the inset nodes were coloured according to the module they represent and their position represents the approximate position of the corresponding modules. Nodes in the inset were labelled with the name of the module centre node in the original network. Note the presence of multiple modules of roughly equal size.

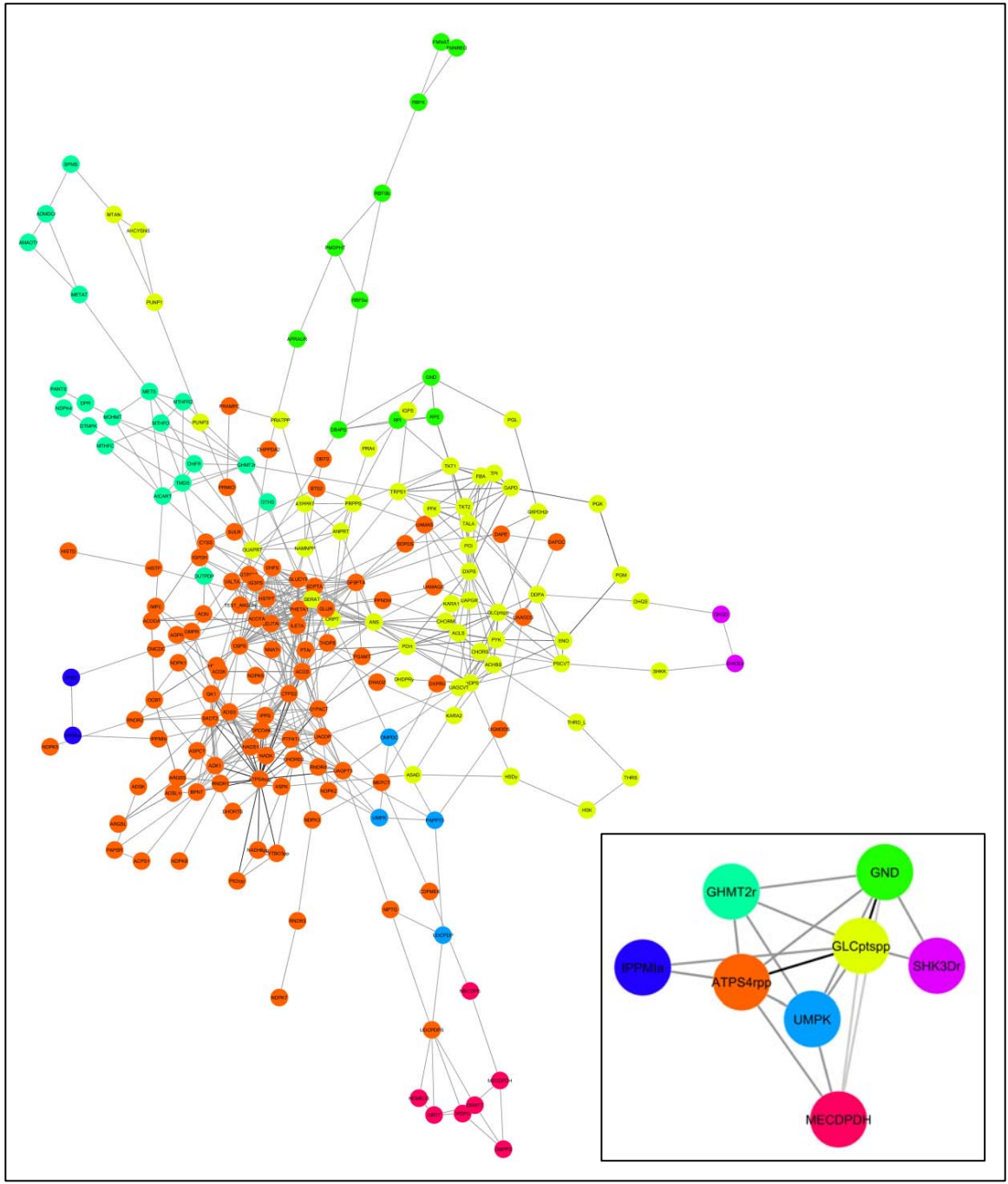

**Supplementary Figure 5. Hierarchical modules of *B. aphidicola* metabolic network.** The construction of the network is described in Supplementary Methods. The inset shows the first level of module hierarchy of the network created by the ModuLand Cytoscape plug-in. Nodes in the inset represent modules of the original network. Images were created by the Cytoscape program [Shannon *et al.*, 2003] in case of both networks, using the Organic yLayout. Edges were coloured in greyscale according to their weights. Colours of the nodes were set by the ModuLand Cytoscape plug-in. Nodes of the original network were coloured according to the colour of the module, where they mostly belong. In the first hierarchical level shown at the inset nodes were coloured according to the module they represent and their position represents the approximate position of the corresponding modules. Nodes in the inset were labelled with the name of the module centre node in the original network. Note the presence of two extremely large modules centred on ATP-synthase (ATPS4rpp, brown) and glucose permease (GLCptspp, yellow).

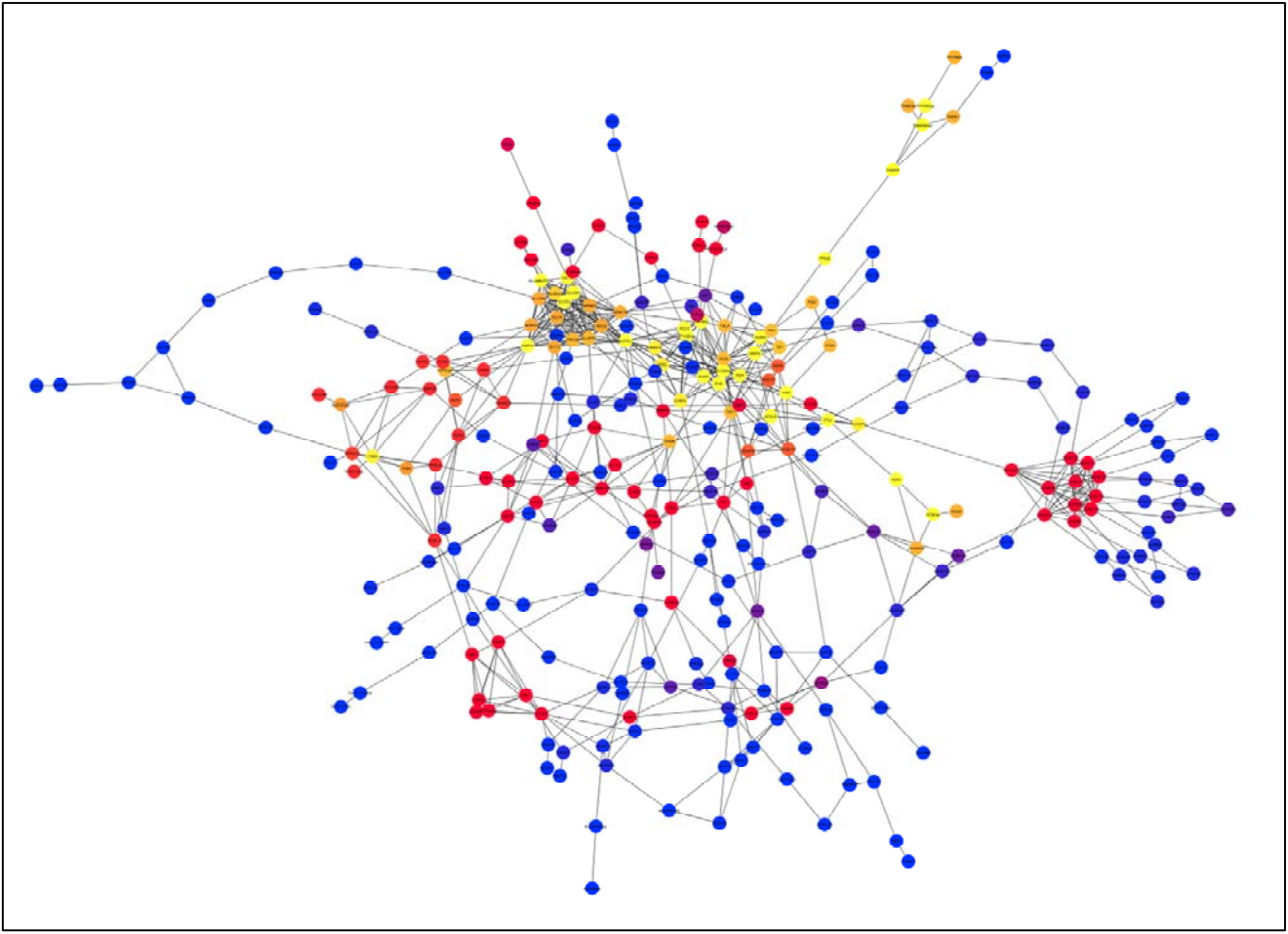

**Supplementary Figure 6. Community centrality landscape of the *E. coli* metabolic network.** The image was created using the Cytoscape program [Shannon *et al*., 2003] with the Organic yLayout. Colours of nodes were set manually by defining custom continuous colour mappings in the Cytoscape program according to the LinkLand community centrality [Kovacs *et al*., 2010] calculated by the ModuLand plug-in. Nodes with community centrality values from 0 to 500 were coloured continuously from blue to red, while the values from 500 to 450,000 were assigned to the colour range between red and yellow. Nodes with community centrality higher than 450,000, were marked with yellow. The largest community centrality value was that of pyruvate kinase (PYK) having a centrality of 2,689,582.

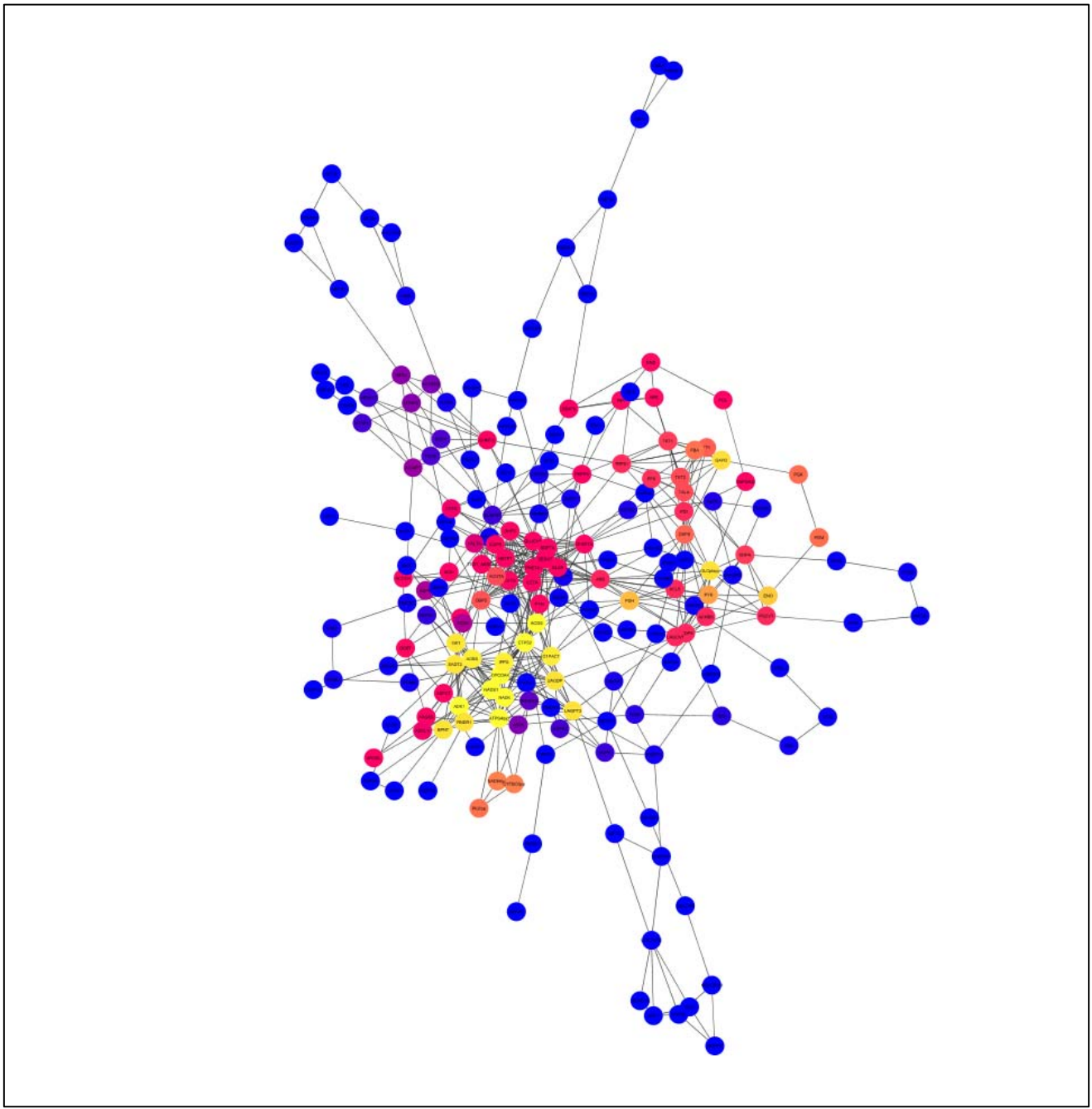

**Supplementary Figure 7. Community centrality landscape of the *B. aphidicola* metabolic network.** The image was created using the Cytoscape program [Shannon *et al.*, 2003] with the Organic yLayout. Colours of nodes were set manually by defining custom continuous colour mappings in the Cytoscape program according to the LinkLand community centrality [Kovacs *et al.*, 2010] calculated by the ModuLand plug-in. Nodes with community centrality values from 0 to 500 were coloured continuously from blue to red, while the values from 500 to 180,000 were assigned to the colour range between red and yellow. Nodes with community centrality higher than 180,000, were marked with yellow. The largest community centrality value was that of ATP synthase (ATPS4rpp) having a centrality of 2,275,416. Note the confluent central plateau in the middle of the network corresponding to the two extremely large modules centred on ATP-synthase and glucose permease (GLCptspp; see Supplementary Figure 5).

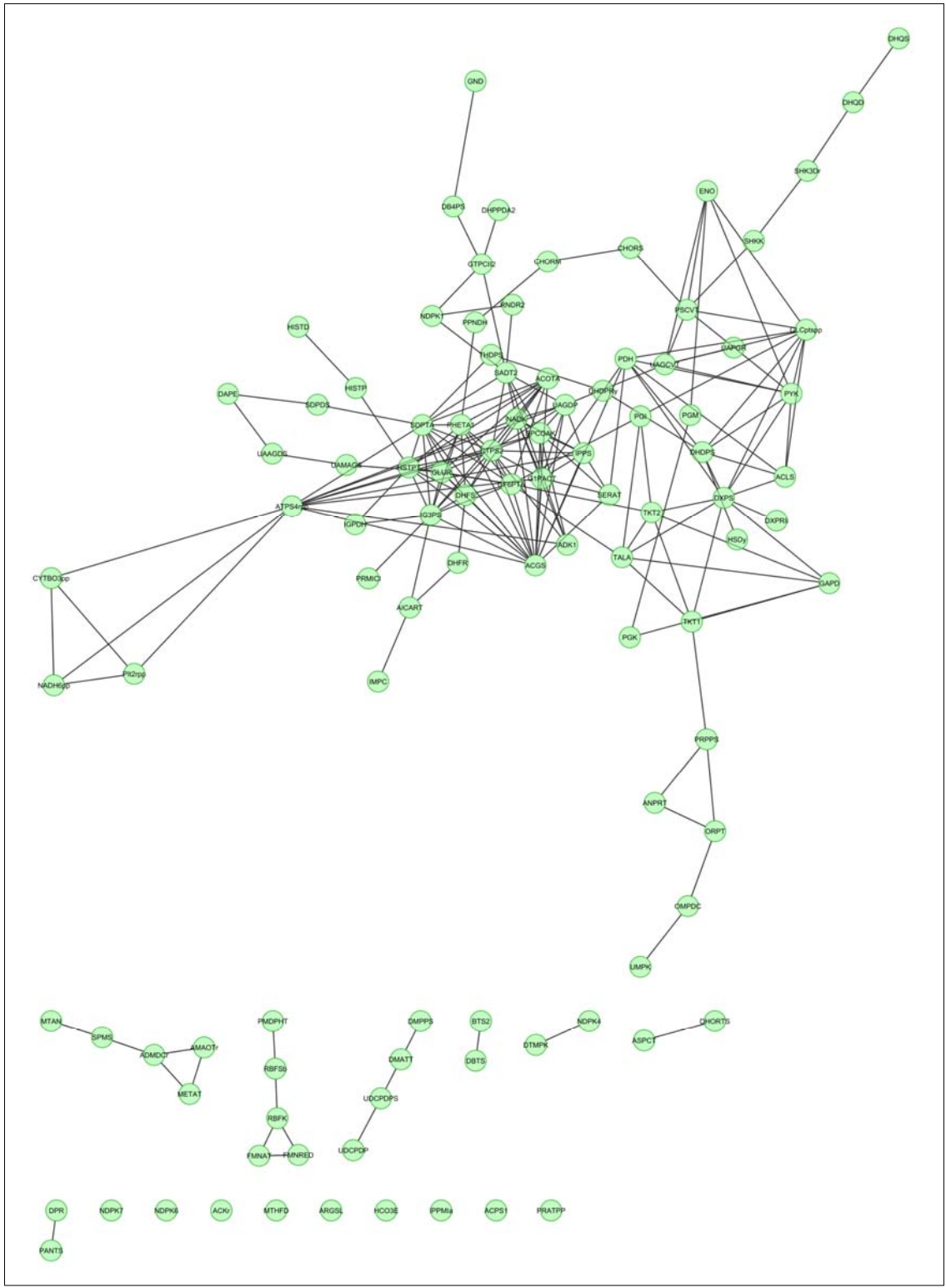

**Supplementary Figure 8. *B. aphidicola* metabolic sub-network containing only the common nodes with the *E. coli* metabolic network**. The network contains the 103 common nodes of the giant components of *B. aphidicola* and *E. coli* metabolic networks connected by their 198 edges from the *B. aphidicola* metabolic network. The image was created using the Cytoscape [Shannon et al., 2003] force directed layout option. The network contains 17 disjoint components, where the largest component has 72 nodes. The list of the common nodes can be found in Supplementary Table 8.

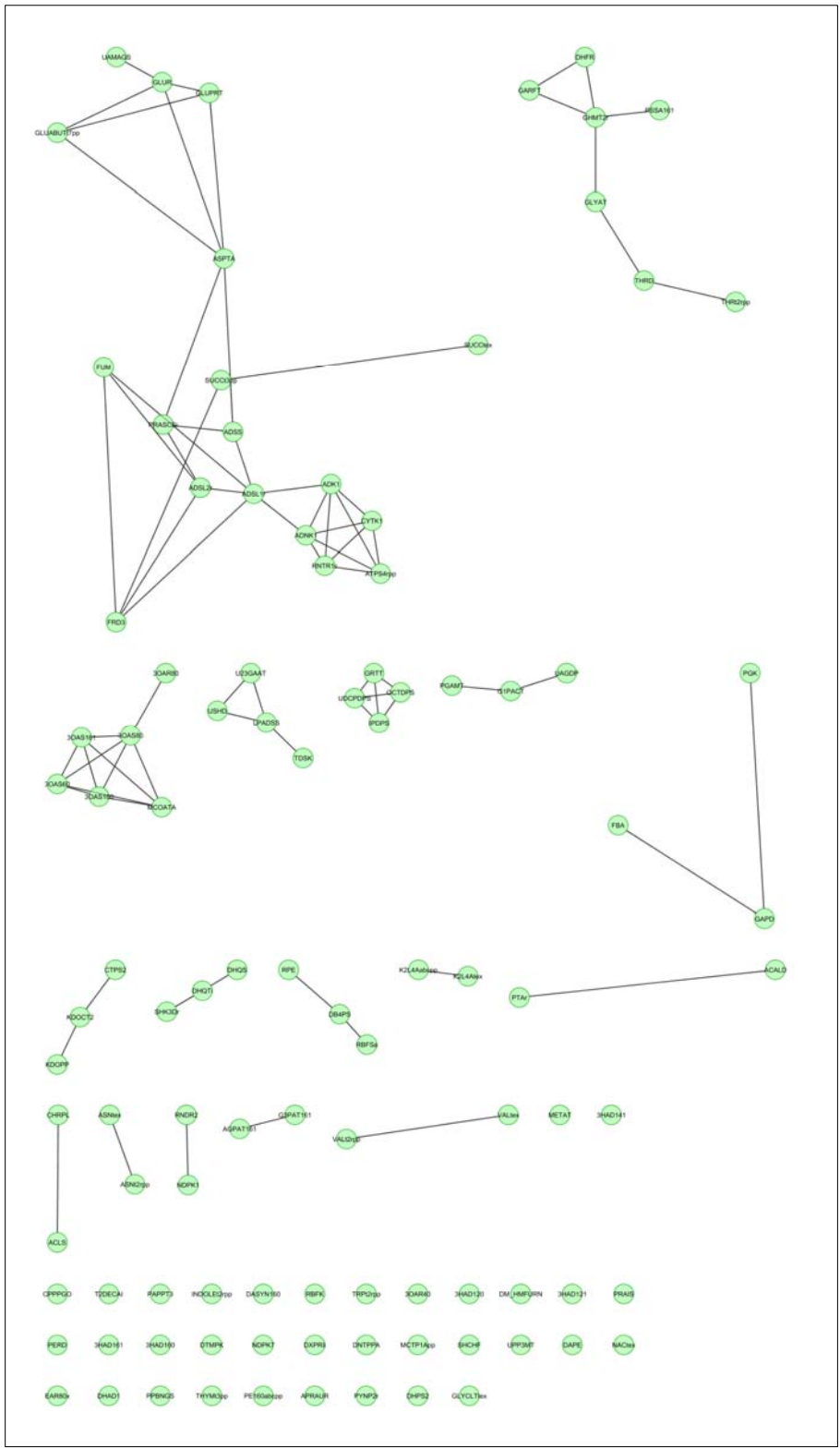

**Supplementary Figure 9. *E. coli* metabolic sub-network containing only the common nodes with the *B. aphidicola* metabolic network.** The network contains the 103 common nodes of the giant components of *B. aphidicola* and *E. coli* metabolic networks connected by their 77 edges from the *E. coli* metabolic network. The image was created using the Cytoscape [Shannon et al., 2003] force-directed layout option. The network contains

52 disjoint components, where the largest component has 18 nodes. The list of the common nodes can be found in Supplementary Table 8.

# Supplementary Tables

**Supplementary Table 1. Domain structure of Met-tRNA synthase.** The table lists the domains of *E. coli* Met-tRNA synthase [Ghosh and Vishveshwara, 2007].

| Domain ID | Amino acid position | Domain name | Subdomain name |
|---|---|---|---|
| 1 | 4-99 | N-terminal, catalytic domain | Rossmann-fold 1 domain (catalytic activity) |
| 2 | 251-322 | | Rossmann-fold 2 domain |
| 3 | 323-388 | | stem contact fold (KMSKS) domain |
| 4 | 100-250 | connecting peptide (CP) domain | |
| 5 | 389-550 | C-terminal, anticodon binding domain | |

**Supplementary Table 2. Modules of Met-tRNA synthase protein structure network.** The table shows the size of modules and the module core amino acids of *E. coli* Met-tRNA synthase protein structure network at the second hierarchical level. Construction of the network is described in Supplementary Methods. Modular structure of the network was determined by the Cytoscape ModuLand plug-in. The first level of hierarchy showed 49 local modules after merging the 47 pairs of the original 96 modules, which were above the 0.9 correlation threshold (not shown). The second hierarchical level detailed in this Table indicated 5 modules corresponding to the domain structure of the protein. Module size was characterized by the 'effective number of amino acids', which efficiently captures the cumulative number of all fractions of amino acids belonging to the given module [Kovacs *et al.*, 2010]. For each module the 10 amino acids having the highest module assignment value of the module (called as module core amino acids) were listed in a decreasing order of modular assignment. Amino acids identified as members of intra-protein communicating pathways [Ghosh and Vishveshwara, 2007] obtained from cross-correlations of molecular dynamics simulations were marked with boldface letters.

| Module ID | Effective number of amino acids | Module core amino acids |
|---|---|---|
| module 1 | 378.0 | Tyr260, **His28**, Trp281, Asn102, Arg356, **Phe84**, Phe264, Gln30, Trp34, Tyr280 |
| module 2 | 103.1 | Trp221, Trp204, Leu201, Asn216, Met218, Glu220, Gln202, Phe197, Phe222, Ser198 |
| module 3 | 178.7 | Tyr490, Tyr418, Met479, Ile400, Arg485, Gln474, Leu473, **Phe484**, Cys477, Ser478 |
| module 4 | 75.8 | **Phe377**, Asn382, Val386, **Asp384**, **Ile385**, Asn387, **Val381**, Val378, Val390, Ala383 |
| module 5 | 65.4 | Cys145, Tyr128, Tyr165, Leu170, Gln126, Ser175, Val141, Lys142, Pro167, Gln153 |

**Supplementary Table 3. Correlations between Met-tRNA synthase domains and protein structure network modules.** Construction of *E. coli* Met-tRNA synthase protein structure network is described in Supplementary Methods. Modular structure of the network was determined by the Cytoscape ModuLand plug-in. Domains of Met-tRNA synthase were assigned as described previously [Ghosh and Vishveshwara, 2007]. Spearman's Rank correlation values were calculated between vectors representing modules and domains by the R statistical program.[3] Each vector had 547 elements, equal to the number of amino acids of *E. coli* Met-tRNA synthase. Values of module vectors were set to the module assignment values of the amino acids at the second level of modular hierarchy as described in the legend of Supplementary Table 2. In case of domain vectors, if the amino acid belonged to the given domain, then its value was set to its community centrality value, while it was zero otherwise. In **Table 3A** correlation of the 5 modules with the 3 major domains are shown. The maximal correlations with the 3 major domains are highlighted with yellow background showing that module 1 corresponds to the catalytic, module 5 to the connecting peptide and module 3 to the anticodon binding domain, respectively. **Table 3B** shows the correlation of the yet un-assigned modules 2 and 4 to the 3 catalytic sub-domains and the 2 other domains of Met-tRNA synthase. Correlation values above 0.2 were highlighted with yellow background showing that module 2 corresponds to both the Rossman-fold 2 subdomain and the adjacent connecting peptide domains, while module 4 corresponds to the stem-contact-fold subdomain and the adjacent anticodon-binding domain.

| Supplementary Table 3A | catalytic domain | connecting peptide domain | anticodon binding domain |
|---|---|---|---|
| module 1 | 0.68 | -0.19 | -0.4 |
| module 2 | 0.33 | 0.28 | -0.58 |
| module 3 | -0.11 | -0.61 | 0.77 |
| module 4 | 0.08 | -0.68 | 0.59 |
| module 5 | -0.24 | 0.8 | -0.53 |

| Supplementary Table 3B | Rossmann-fold 1 (catalytic) subdomain | Rossmann-fold 2 subdomain | stem contact fold subdomain | connecting peptide domain | anticodon binding domain |
|---|---|---|---|---|---|
| module 2 | 0.13 | 0.22 | 0.13 | 0.28 | -0.58 |
| module 4 | -0.02 | -0.27 | 0.42 | -0.68 | 0.59 |

---

[3] R is a GNU project defines a language and gives an environment for statistical computing and graphics. See: http://www.r-project.org

**Supplementary Table 4. Modular properties of the shortest and most frequently used Met-tRNA synthase communication pathway.** The Table contains amino acids of pathway IV of Ghosh and Vishveshwara [2007] obtained from cross-correlations of molecular dynamics simulations. Pathway IV starts at Leu13 of the catalytic centre, and propagates the conformational change towards Trp461, which constitutes a part of the tRNA anticodon binding site. Domain numbers refer to those of Supplementary Table 1. Module assignment values show the strength of the assignment of each amino acid to different modules. Values higher than 30% are highlighted violet for each amino acid. Community centrality, overlap and bridgeness values (defined as in [Kovacs *et al.*, 2010]) were calculated by ModuLand Cytoscape plug-in. On the table we highlighted those numbers yellow, which belonged to the top 15%.

| Communicating amino acid | Domain | Module assignment ratios (%) | | | | | Community centrality | Overlap | Bridgeness |
|---|---|---|---|---|---|---|---|---|---|
| | | module 1 (catal. dom.) | module 2 (catal. + conn. pept. dom.) | module 3 (anti-codon dom.) | module 4 (catal. + anti-codon dom.) | module 5 (conn. pept. dom.) | | | |
| Leu13 | 1 | 98.31 | 0.56 | 1.09 | 0.03 | 0.00 | 858 | 1.10 | 0.0521 |
| His28 | 1 | 95.51 | 0.39 | 3.98 | 0.12 | 0 | 6653 | 1.22 | 0.0683 |
| Ile89 | 1 | 95.42 | 0.39 | 4.07 | 0.12 | 0 | 2710 | 1.23 | 0.0197 |
| Asp32 | 1 | 95.42 | 0.39 | 4.07 | 0.12 | 0 | 6532 | 1.23 | 0.0549 |
| Arg36 | 1 | 93.73 | 0.38 | 5.66 | 0.23 | 0 | 6327 | 1.29 | 0.0755 |
| Leu495 | 5 | 36.81 | 0.04 | 55.30 | 7.85 | 0 | 2465 | 2.46 | 0.0626 |
| Tyr357 | 3 | 57.16 | 0.34 | 30.37 | 12.13 | 0 | 1268 | 2.60 | 0.0682 |
| Asp384 | 3 | 8.32 | 0.06 | 15.51 | 76.11 | 0 | 2468 | 2.03 | 0.0667 |
| Lys388 | 3 | 7.05 | 0.03 | 29.63 | 63.29 | 0 | 3207 | 2.31 | 0.1200 |
| Asn452 | 4 | 13.51 | 0.01 | 79.59 | 6.89 | 0 | 4186 | 1.89 | 0.0881 |
| Arg395 | 4 | 8.48 | 0.01 | 66.40 | 25.11 | 0 | 1138 | 2.29 | 0.0357 |
| Asp456 | 4 | 13.97 | 0.01 | 84.52 | 1.51 | 0 | 3403 | 1.62 | 0.0172 |
| Trp461 | 4 | 13.87 | 0.01 | 84.58 | 1.55 | 0 | 905 | 1.62 | 0.0047 |

**Supplementary Table 5. Community centrality, overlap and bridgeness values of Met-tRNA synthase communicating amino acids.** The table shows the average values of community centrality, overlap and bridgeness values calculated by the ModuLand Cytoscape plug-in based on the second hierarchical level as described in Supplementary Table 2. Communicating amino acids denote the 43 amino acids defined as members of communicating pathways between the active centre and the anticodon binding site by [Ghosh and Vishveshwara, 2007] obtained from cross-correlations of molecular dynamics simulations. Communicating amino acids have higher average values than the rest of the network in case of all the three measures. The difference between the two datasets was verified using the Welch's two sample t-test (p-value < 0.0008 for all the three measurements).

| | Number of amino acids | Community centrality | Overlap | Bridgeness |
|---|---|---|---|---|
| Communicating amino acids | 43 | 2895 | 1.86 | 0.065 |
| Not communicating | 504 | 1795 | 1.50 | 0.034 |

| **amino acids** | | | |

**Supplementary Table 6. Basic structural properties of *E. coli* and *B. aphidicola* metabolic networks**. Values were calculated by the Python igraph module [Csardi and Nepusz, 2006]. The characteristic path length was defined as the average length of the unweighted shortest paths, while the diameter of the network was the length of the longest unweighted path containing no circles. The global clustering coefficient (transitivity) was defined as the probability that the neighbours of a node are connected. We created random samples from the *E. coli* network by selecting connected random sub-networks with the same number of nodes or edges as can be found in the *B. aphidicola* network. In each case 1000 random sample sub-networks of the *E. coli* metabolic network having an equal number of nodes or edges like the *B. aphidicola* network were selected as described in the Methods section in detail. Table shows average values ± the standard deviations.

| Organism | *E. coli* | *E. coli* samples (node limit = 190 as in *B. aphidicola*) | *E. coli* samples (edge limit = 563 as in *B. aphidicola*) | *B. aphidicola* |
|---|---|---|---|---|
| **Number of nodes** | 294 | 190 ±0 | 258 ±5.4 | 190 |
| **Number of edges** | 730 | 328 ±22.9 | 559 ±4.04 | 563 |
| **Global clustering coefficient** | 0.54 | 0.58 ±0.04 | 0.59 ±0.02 | 0.6 |
| **Average number of neighbours** | 4.966 | 3.45 ±0.24 | 4.34 ±0.1 | 5.93 |
| **Characteristic path length** | 5.74 | 8.86 ±1.16 | 6.56 ±0.35 | 4.20 |
| **Diameter** | 15 | 24.61 ±3.95 | 18.13 ±2.2 | 11 |

**Supplementary Table 7. Modular properties of *E. coli* and *B. aphidicola* metabolic networks.** Values were calculated on a Linux machine using the same ModuLand binary modularization programs which are packaged to the Cytoscape plug-in. We generated two types of random samples from the *E. coli* metabolic network. In each case we selected 1000 random sample sub-networks of the *E. coli* metabolic network having an equal number of nodes or edges like the *B. aphidicola* network as described in the Methods section in detail. The table shows the average values ± the standard deviations. Average module size values were the ratios between node and module numbers, while the average effective module sizes were the average values of the effective number of nodes belonging to the given module. The effective number of modules was calculated based on the module assignment values summed up for each node. The exact definitions of the metrics are described in the Supplementary Information of [Kovacs *et al.*, 2010].

| Organism | *E. coli* | *E. coli* samples (node limit = 190 as in *B. aphidicola*) | *E. coli* samples (edge limit = 563 as in *B. aphidicola*) | *B. aphidicola* |
|---|---|---|---|---|
| **Number of nodes** | 294 | 190 ±0 | 258 ±5.4 | 190 |
| **Number of edges** | 730 | 328 ±22.9 | 559 ±4.04 | 563 |
| **Number of modules** | 23 | 21.97 ±2.67 | 23.66 ±2.4 | 8 |
| **Effective number of modules** | 6.16 | 5.5 ±0.95 | 5.61 ±0.63 | 2.11 |
| **Average module size** | 12.78 | 8.78 ±1.1 | 11 ±1.1 | 23.75 |
| **Average effective module size** | 8.07 | 5.81 ±0.46 | 7.14 ±0.49 | 11.34 |
| **Average module overlap of nodes** | 1.69 | 1.49 ±0.1 | 1.28 ±0.08 | 1.4 |

**Supplementary Table 8. The common metabolites of *E. coli* and *B. aphidicola* metabolic networks.** We list the 103 common nodes of the giant components of *E. coli* and *B. aphidicola* metabolic networks. Names of the nodes are the same in case of 102 nodes, while there is one additional metabolite (named DHQD in the *B. aphidicola* network and DHQTi in the *E. coli* network) with different names.

| Name of the metabolite | | | |
|---|---|---|---|
| ACKr | GF6PTA | PMDPHT |
| ACLS | GHMT2r | PRPPS |
| ADK1 | GK1 | PSCVT |
| ADSL1r | GLCptspp | PTAr |
| ADSS | GLUR | PUNP1 |
| AHCYSNS | GRTT | PYK |
| AICART | GTPCII2 | RBFK |
| APRAUR | HCO3E | RBFSa |
| ASAD | IMPC | RBFSb |
| ASPK | IPDPS | RNDR2 |
| ATPS4rpp | KARA1 | RNDR3 |
| CDPMEK | MECDPS | RPE |
| CHORS | MEPCT | RPI |
| CTPS2 | METAT | SDPDS |
| DAPE | METS | SDPTA |
| DB4PS | MPTG | SHK3Dr |
| DDPA | MTHFC | SHKK |
| DHDPRy | MTHFD | TALA |
| DHDPS | MTHFR2 | THDPS |
| DHFR | NADK | TKT1 |
| DHFS | NADS1 | TKT2 |
| DHPPDA2 | NAMNPP | TPI |
| DHQD (DHQTi) | NDPK1 | UAAGDS |
| DHQS | NDPK2 | UAGCVT |
| DMATT | NDPK3 | UAGDP |
| DMPPS | NDPK4 | UAGPT3 |
| DTMPK | NDPK5 | UAMAGS |
| DXPRIi | NDPK7 | UAMAS |
| DXPS | NNATr | UAPGR |
| ENO | PAPPT3 | UDCPDP |
| FBA | PDH | UDCPDPS |
| FMNAT | PGAMT | UGMDDS |
| G1PACT | PGI | UMPK |
| GAPD | PGK | VALTA |
| | PGM | |

**Supplementary Table 9. Comparing the various clustering plug-ins available for Cytoscape.** Numerous clustering methods are available as Cytoscape plug-ins. The following table compares the most widely used clustering Cytoscape plug-ins with the ModuLand Cytoscape plug-in. References and more information for each plug-in can be found in the Supplementary Discussion.

| Plug-in name | Clustering method(s) | Cytoscape baseline for the latest plug-in | Module overlaps | Supported platforms* | Additional feature(s) |
|---|---|---|---|---|---|
| ModuLand | Overlapping modules are determined based on node centrality/density values defined by limited network walks started from each edge | 2.8.2 | yes | any | Determining overlapping module hierarchy, calculating measures based on overlapping module structure, colouring of the network, etc. |
| MCODE | Finds clusters based on local density measures | 2.5.1 | no | any | Creating networks from the identified clusters; fine-tuning the selected cluster's size |
| MINE | Agglomerative clustering algorithm (based on local density and modularity measures) | 2.6 | yes | any | Creating networks from the identified clusters |
| NeMo | Identifies modules based on a neighbour-sharing score | 2.7 | no | any | |
| clusterMaker | Contains various clustering methods (e.g. hierarchical, k-means, AutoSOME, affinity propagation, MCODE, etc.) | 2.8.2 | no | any | Cluster visualizations (e.g. tree view, heat map); creating new network from clusters or attributes |
| GLay | Clauset-Newman-Moore method implementation variants | 2.7.0 | no | any | Including layout algorithms |
| | Clustering methods from igraph library | | | windows 64bit | |

*: 'any' platform includes all platforms supported by Cytoscape (like Windows, Linux, Mac OS; both 32 and 64bit)

**Supplementary Table 10. Runtime of the ModuLand plug-in in case of different networks.** Network modularization and the creation of the first hierarchical module level were performed using the ModuLand Cytoscape plug-in. The runtime was measured both with and without the optional optimization included in the ModuLand plug-in for large networks. (This optimization reduces the number of low intensity edges appearing in higher hierarchical levels reflecting the minor overlaps between distant modules at one level lower in the hierarchy.) All values listed below show the average time of five runs. All modularizations were run on the same software and hardware environment: Intel Core2 Duo 3 GHz processor, 4GB RAM, 32bit Windows 7 Professional, Cytoscape 2.8.2, Oracle Java SE 1.6.26. The *B. aphidicola* and *E. coli* metabolic networks were created using the data of [Feist *et al.*, 2007] and [Thomas *et al.*, 2009]. The *E. coli* Met-tRNA synthase protein structure network is described in [Ghosh and Vishveshwara, 2007]. The yeast high fidelity protein interaction network was assembled by [Ekman *et al.*, 2006]. The school friendship network was constructed based on [Moody, 2001]. The power distribution network was defined in [Watts and Strogatz, 1998]. The word association network is based on the University of South Florida word association network (http://www.usf.edu/FreeAssociation). All the seven networks are further described in the Supplementary Discussion and can be downloaded from http://www.linkgroup.hu/modules.php.

| Network name | Number of nodes | Number of edges | Number of modules | Runtime [sec] without optimization | Runtime [sec] with optimization |
|---|---|---|---|---|---|
| *B. aphidicola* metabolic network | 190 | 563 | 8 | 0.82 | 0.78 |
| *E. coli* metabolic network | 294 | 730 | 23 | 0.92 | 0.84 |
| *E. coli* Met-tRNA synthase protein structure network | 547 | 2153 | 96 | 2.19 | 2.04 |
| Yeast high fidelity interactome | 2444 | 6271 | 55 | 9.60 | 8.91 |
| School friendship network | 1127 | 5096 | 236 | 14.91 | 8.88 |
| Power distribution network | 4941 | 6594 | 207 | 23.61 | 22.98 |
| Word association network | 10617 | 63788 | 994 | 4935.51 | 702.97 |

# Supplementary Methods, Results and Discussion

**Construction of *E. coli* Met-tRNA synthase protein structure network**

The protein structure network of *Escherichia coli* Met-tRNA syntethase was generated from the equilibrated state of the molecular simulation of the *E. coli* Met-tRNA synthase/tRNA/MetAMP complex as described and kindly shared by [Ghosh and Vishveshwara, 2007]. The network was obtained by converting the Cartesian coordinates of the 3D image to distances of amino acid pairs, and keeping all non-covalently bonded contacts within a distance of 0.4 nm. The final weighted network was created by removing self-loops, and calculating the inverse of the average distance between amino acid residues as edge weights. The protein structure network had 547 nodes and 2,153 edges, since the first 3 N-terminal amino acids were not participating in the network. The network data can be downloaded from our web-site: <www.linkgroup.hu/modules.php>.

**Construction of *E. coli* and *B. aphidicola* metabolic networks and randomly selected sub-networks of the *E. coli* metabolic network**

Metabolic networks of *Escherichia coli* and *Buchnera aphidicola* were constructed based on the primary data of [Feist *et al.*, 2007] and [Thomas *et al.*, 2009], respectively. Frequent cofactors were deleted from the networks, except of those metabolic reactions, where cofactors were considered as main components. For better comparison of networks, metabolic reactions were assumed to be irreversible and flux balance analyses (FBA) were performed resulting in weighted networks. All flux quantities were minimized, whereas reactions not affecting the biomass production were considered having zero flux. Weights were generated as the mean of the appropriate flux quantities in absolute value, except of the case when one of the fluxes was zero that resulted in a zero weight automatically. Subnetworks were created based on metabolic reactions having non-zero flux quantities, and the giant components of the respective networks were analyzed using the ModuLand Cytoscape plug-in. Metabolic networks of *B. aphidicola* or *E. coli* had 190 nodes and 563 edges, or 294 nodes and 730 edges, respectively. The same networks were used earlier [Mihalik and Csermely, 2011]. The network data can be downloaded from our web-site: <www.linkgroup.hu/modules.php>.

We selected connected random sub-networks from the *E. coli* metabolic network using the algorithm having the following pseudo code:

```
repeat {
        sub-network := original network
        while the sub-network has more node (or edge) than the node (or edge) limit {
                repeat {
                        rnd := a randomly choosed node from the sub-network
                        tmp_network := sub-network – rnd
                } until tmp_network has only one component
                sub-network := sub-network – rnd
        }
        if the sub-network is different from each network in the storage then store the sub-network
} until the storage does not contain enough sub-networks
```

The two ensembles of 1000 randomly selected sub-networks and the Python scripts generating the sub-networks can be downloaded from our web-site: <www.linkgroup.hu/modules.php>.

**Construction of a school-friendship network**

We used the data of the high-scale Add-Health survey, which mapped social connections of high schools of the USA [Gonzaléz et al., 2007; Moody, 2001, Newman, 2003].[4] In the survey recorded between 1994 and 1995 social connections of 90,118 students in 84 schools were recorded. For each friend named, the student was asked to check off, whether he/she participated in any of five activities with the friend. These activities were:

1. you went to (his/her) house in the last seven days;
2. you met (him/her) after school to hang out or go somewhere in the last seven days;
3. you spent time with (him/her) last weekend;
4. you talked with (him/her) about a problem in the last seven days;
5. you talked with (him/her) on the telephone in the last seven days.

Based on these data, connections were assigned with weights from 1 to 6. A nomination as friend already resulted in a weight of one, and each checked category added one to that weight. In addition to the nomination data, these files include the gender, race, grade in school, school code, and total number of nominations made by each student.

We measured the runtime of the ModuLand plug-in using the Community-44 school network, because it contains a high number of students with a dense social network [Newman, 2003]. This network has an approximately equal number of black and white students. The network contains 1,147 students with 6,189 directed edges between them. In our current study directed parallel edges were merged into a single undirected edge with a weight equal to the sum of the original weights, and only giant component of the network was used. This process resulted in a weighted undirected network consisting of 1,127 nodes and 5,096 edges with weights between 1 and 12. The network data can be downloaded from our web-site: <www.linkgroup.hu/modules.php>.

**Construction of the electrical power-grid network of the USA**

To measure the runtime of the ModuLand plug-in, we used the unweighted and undirected USA Western Power Grid network as an example from the field of engineered networks [Watts and Strogatz, 1998]. The power grid network has 4,941 nodes and 6,594 edges, and is a favored network for studying error propagation and the effect of malicious attacks. The original network data were downloaded from the website of Prof. Duncan Watts (University of Columbia, http://cdg.columbia.edu/cdg/datasets). The network data can also be downloaded from our web-site: <www.linkgroup.hu/modules.php>.

**Construction of the yeast protein-protein interaction network**

To measure the runtime of the ModuLand plug-in, we used the unweighted and undirected yeast protein-protein interaction network assembled by Ekman et al. [2006] consisting of 2,633 nodes and 6,379 edges covering approximately half the proteins of yeast genome. We analyzed the largest connected component of the network consisting of 2,444 nodes and 6,271 edges. Besides the high confidence of its data, we chose this network, because it was used in the identification of party and date hubs, an interesting dynamic feature of protein-protein interaction networks [Ekman et al., 2006]. The network data can be downloaded from our web-site: <www.linkgroup.hu/modules.php>.

---

[4]This research uses data from Add Health, a program project designed by J. Richard Udry, Peter S. Bearman, and Kathleen Mullan Harris, and funded by a grant P01-HD31921 from the National Institute of Child Health and Human Development, with cooperative funding from 17 other agencies. Special acknowledgment is due Ronald R. Rindfuss and Barbara Entwisle for assistance in the original design. Persons interested in obtaining data files from Add Health should contact Add Health, Carolina Population Center, 123 W. Franklin Street, Chapel Hill, NC 27516-2524 (addhealth@unc.edu).

**Construction of a word association network**

To measure the runtime of the ModuLand plug-in, we used the University of South Florida word association network (http://www.usf.edu/FreeAssociation), where 6,000 participants produced nearly three-quarters of a million responses to 5,019 stimulus words. This word association network gives a relative strength for each stimulus-response word pair, calculated by taking into consideration the count of associations to the response word given the count of stimuli by the stimulus word: The relative weight of an A -> B edge (called forward strength, FSG) is expressed as FSG = P/G, where G is the count of people who received the word A as the stimulus, and P is the count of people among them who responded with word B for that stimulus. Based on this data, a weighted and directed network can be built. While the direction of edges provide insight to the complexity of human conceptual thinking, in the present study we considered the fact of association between words, and built an undirected network. Therefore, the parallel forward and backward edges were collapsed into a single non-directed edge, and weighted with the sum of the original weights. This process on the giant component of Appendix A of the University of South Florida word association network resulted in a weighted and undirected network. In this study we analyzed the largest connected component of this network consisting of 10,617 nodes (English words) and 63,788 edges (associations) between them. The network data can be downloaded from our web-site: <www.linkgroup.hu/modules.php>.

**Correlations between Met-tRNA synthase domains and protein structure network modules**

Clustering analysis has been used for a long time to identify protein domains [Guo *et al*., 2003; Xu *et al*., 2000]. However, former methods used the non-overlapping modularization technique based on the classical minimum-cut *Ford-Fulkerson* algorithm. This method performs well with two-domain proteins, but gives not so precise results with multi-domain proteins [Xu *et al*., 2000]. Later the two-domain cut algorithm was extended by a neural network learning mechanism [Guo *et al*., 2003]. The difficulties of domain prediction made interesting to examine whether the high-resolution ModuLand algorithm may predict the domains of a larger protein.

The *E. coli* Met-tRNA synthase enzyme contains 550 amino acids forming 3 major domains, the catalytic, the tRNA-binding and the connecting domains. Its catalytic domain is consisted of 3 sub-domains: 2 Rossmann-folds and a stem-contact fold. The first Rossmann-fold contains the active centre of the enzyme (Supplementary Table 1. and [Ghosh and Vishveshwara, 2007]). The modular structure of the first hierarchical level is shown on Panel B of Supplementary Figure 1. This level has the 547 amino acids of the protein structure network and their 3D physical contacts as 2,153 weighted edges. Panel C of Supplementary Figure 1 shows the second hierarchical level, where the 49 local modules of the first level serve as nodes of the second level and their 490 overlaps give 490 weighted edges of the second hierarchical level. At this, second level 5 modules have been identified as shown on Supplementary Table 2.

The effective number of modules at the second hierarchical level is *3.2*, which is roughly the same as the number of the 3 major domains. Supplementary Table 3A lists the Spearman's Rank correlation of the 3 major domains with the 5 modules at this level of hierarchy. The N-terminal, catalytic domain, the connecting peptide domain and the anticodon binding domain are mostly correlating with modules 1, 3 and 5, respectively. As shown on Supplementary Table 3B module 2 corresponds mainly to both the Rossmann-fold 2 subdomain and the connecting peptide domain. Module 4 corresponds to the stem contact fold subdomain and the anticodon binding domain. Importantly, both the {Rossmann-fold 2/connecting peptide} and the {stem contact fold/anticodon binding domain} pairs are adjacent to each other in the primary structure of the protein. The fact, that only modules at higher hierarchical levels correspond to the domains of the protein, is similar to the findings of Delvenne *et al.*, [2010] and Delmotte *et al*. [2011], who described a large number of smaller, initial modules, which merged to larger clusters corresponding to the domain structure of the proteins examined only at the end of the simulation. The modular structure obtained here is in agreement of the structure obtained before for the same enzyme by Ghosh and Vishveshwara [2008] using the overlapping modularization program, CFinder [Adamcsek *et al.*, 2006]. This former analysis found disjoint modules, which is in agreement of the very cohesive nature of the modules found by the CFinder program (for a direct comparison, see [Kovacs *et al.*, 2010].

**Modular properties of Met-tRNA synthase communicating amino acids**
Met-tRNA synthase needs to recognize both the anticodon and aminoacylation regions of the tRNA, which are relatively far from each other (separated by ~70Å in space). An earlier study [Ghosh and Vishveshwara, 2007] examining the cross-correlations of molecular dynamic simulations of the protein combined with a continuity analysis of the protein structure network identified four communication pathways of 43 amino acids accomplishing the propagation of conformational changes during the process of aminoacylation. Since the module core amino acids listed in Supplementary Table 2 have the largest module assignment value of their module, which often corresponds with high community centrality values (community centrality being the sum of all modular assignment values of the given node [Kovacs et al., 2010]), these core amino acids often have the highest influence on their own module. Thus the involvement of these module core amino acids may enhance the robustness of intra-protein signal transduction. Importantly, 7 communicating amino acids are module core amino acids of modules 1, 3 and 4, which are the modules participating in the intra-protein signalling events (Supplementary Tables 2 and 4). Communicating amino acids of the module cores are key factors in communication pathways II, III and IV. Interestingly, the only intra-protein signalling pathway not represented in module cores, pathway I, is the least frequently used pathway, which was active only in 2.3% of the simulations in the original publication [Ghosh and Vishveshwara, 2007].

Nodes of the module cores are generally more influential to determine the module function and intra-domain communication than the rest of the modules. This feature of module cores has also been shown in our other studies on protein-protein interaction networks [Mihalik and Csermely, 2011] as well as on chromatin networks (Sandhu, K.S., Li, G., Poh, H.M., Quek, Y.L.K., Sia, Y.Y., Peh, S.Q., Mulawadi, F.H., Sikic, M., Menghi, F., Thalamuthu, A., Sung, W.K., Ruan, X., Fullwood, M.J., Liu, E., Csermely, P. and Ruan, Y. Large scale functional organization of long-range chromatin interaction networks, submitted for publication). In our earlier work [Mihalik and Csermely, 2011] module cores were often referred as nodes having the highest community centrality in the module. In this earlier work we used the NodeLand version of the ModuLand method, which makes a much closer correlation between the highest intra-modular community centrality values and largest module assignment values than the LinkLand version of the ModuLand method used in the plug-in and in the current paper.

Since the two most important segments of Met-tRNA synthase recognizing the anticodon and aminoacylation tRNA regions are in different modules, their communication must involve inter-modular regions besides the module cores identified above. Two measures of the ModuLand method [Kovacs *et al.*, 2010], the overlap and the bridgeness mark different inter-modular positions. Overlap measures the effective number of modules, where the given amino acid belongs. This measure is close to 1, if the amino acid is a part of a module core, since in such cases the amino acid essentially belongs to a single module. The overlap value is increasing above 1 for amino acids situated equally close to different module cores. The bridgeness value involves the smaller of the two modular assignments of an amino acid in two adjacent modules. To give the bridgeness value, these smaller values of the modular assignment-pairs are summed up for every module pairs. This value is high, if the amino acid belongs more equally to two adjacent modules in many cases, *i.e.* if it behaves as a bridge between a single pair, or between multiple pairs of modules. Such bridging positions correspond to saddles between the 'community-hills' of the 3D community landscape shown on Supplementary Figure 1. Note that the bridgeness measure characterizes an inter-modular position of the amino acid between adjacent modules, while the overlap measure reveals the simultaneous involvement of the amino acid in multiple modules.

The amino acid having the highest overlap value is Tyr531, which is connecting communication pathways II and III [Ghosh and Vishveshwara, 2007] as well as modules 1, 2 and 4. Two among the top 15 bridge amino acids are identified as members of intra-protein pathways. Leu392 has the $7^{th}$ highest bridgeness and the $23^{rd}$ highest overlap value (top 5%) in the whole network. Leu392 is part of the anticodon binding domain, and bridges modules 3 and 4 correlating with this domain. Communication pathways I, II and III all go trough Leu392 as their bridging amino acid [Ghosh and Vishveshwara, 2007]. Trp432 has also a top bridging position ($13^{th}$, top 2.5%). The high community centrality value ($43^{rd}$ highest, top 8%) of Trp432 shows that this amino acid is located in a dense part of the network. Trp432 bridges modules 1 and 3, was identified as a key member of communication pathway I, and serves as an interface between the catalytic and anticodon binding domains [Ghosh and Vishveshwara, 2007].

Supplementary Table 4 shows the shortest communication pathway, pathway IV, between the anticodon region and the active centre of Met-tRNA synthase. Pathway IV was the most frequently used pathway participating in intra-protein signalling processes in 43.3% of simulations in the original publication [Ghosh and Vishveshwara, 2007]. Pathway IV starts from His28, which is the $2^{nd}$ most central amino acid of module 1 corresponding to the catalytic domain. The continuation of pathway IV, Ile89, Asp32 and Asp36 also form a part of module 1. The middle segment of pathway IV propagates through Leu495, Tyr357, Asp384 and Lys388, which are at an overlapping region between modules 1, 3 and 4 corresponding to the stem-contact fold subdomain and the anticodon binding domain. Pathway IV converges with other communication pathways at amino acids Asn452, Arg395, Asp456 and Trp461, which are all belonging to module 3 corresponding to the anticodon binding domain.

The 43 amino acids of the four pathways transmitting the conformational changes from the catalytic centre to the anticodon binding region of Met-tRNA synthase [Ghosh and Vishveshwara, 2007] all have higher average community centrality, overlap and bridgeness values than the rest of the protein (Supplementary Figure 3 and Supplementary Table 5). The relatively high deviation makes the prediction of communicating amino acids from modular data rather difficult, but the results clearly indicate the preference of intra-protein signalling for central, overlapping or bridge-like protein regions, which in agreement with general assumptions [Csermely *et al.*, 2012; Farkas *et al.*, 2011] and earlier findings [Del Sol *et al.*, 2007; Ghosh and Vishveshwara, 2008]. It is of particular note, that Ghosh and Vishveshwara [2008] found the structure analyzed in this paper the most flexible structure of the enzyme. Modular analysis of the communicating amino acids of this structure showed the participation of two separate clique structures, which indicates that the current modularization method offers a more detailed picture of highly mobile, flexible systems, than other methods. Communication pathways in other tRNA synthases, like the Glu- and Leu-tRNA synthase from various bacteria show a similar pattern using several alternative pathways for transmission and showing a convergence of the transmission pathways at critical nodes of inter-modular boundaries [Sethi *et al.*, 2009]. Moreover, the large overlap between highly mobile and inter-modular network nodes is similar to our earlier finding using protein-protein interaction networks, where inter-modular nodes, having both a large bridgeness and community centrality at the same time, corresponded to date hubs [Kovacs *et al.*, 2010]. Recent studies also uncovered the usefulness of complex centrality measures, similar to the community centrality used here, in the identification of biologically important network nodes [Milenkovic *et al.*, 2011].

As a summary of our studies, we may say that network communication in general (and the transmission of allosteric signals in particular) may preferentially involve two types of nodes: 1.) intra-modular nodes forming a module core (often having a high community centrality, i.e. high communication level with the rest of the module); and 2.) inter-modular nodes (either bridges preferentially connecting 2 modules or overlapping nodes connecting more modules at the same time). Signals may often propagate using and alternating sequence of module cores and inter-modular nodes [see also Csermely et al., 2012].

**Structural properties and modular analysis of *E. coli* and *B. aphidicola* metabolic networks**

Supplementary Figure 2 illustrates a few key centrality measures of the ModuLand Cytoscape plug-in (such as community centrality, bridgeness and betweenness centrality) on the *Escherichia coli* metabolic network. Supplementary Table 6 shows the basic structural properties of the metabolic networks of *Escherichia coli* and *Buchnera aphidicola*. The number of nodes in the *E. coli* network is 54% larger than that of the *B. aphidicola* metabolic network, but the *E. coli* metabolic network has relatively less edges, since the average number of neighbours is 5 and 6 in the *E. coli* and *B. aphidicola* networks, respectively.

To rule out the effects of the different network size in the comparison of network topology measures of *E. coli* and *B. aphidicola* metabolic networks, we created random samples from the larger *E. coli* network by selecting connected random sub-networks with the same number of nodes or edges that can be found in the *B. aphidicola* network as described in the Methods section of this Supplement in detail. In each case we selected 1000 random sample sub-networks of the *E. coli* metabolic network having an equal number of nodes or edges like the *B. aphidicola* network, and calculated the average values ± the standard deviations of network topology measures. Results are summarized in Supplementary Table 6. Both the characteristic path length and the network diameter were higher in the *E. coli* than in *B. aphidicola,* and became even higher both in the node-limited and edge-limited random sample sub-networks of the *E. coli* metabolic network.

The above differences together already suggest a multi-centred network structure of the *E. coli* metabolic network as compared to a more centralized network of *B. aphidicola*. Such a difference is also quite prevalent by the visual inspection of the two networks, where the modular structure is multifocal in case of *E. coli*, while it is dominated by the twin super-modules of ATP-synthase and D-glucose transport in case of *B. aphidicola* (cf. Supplementary Figures 4 and 5). The community centrality landscape shows a similar pattern having multiple groups of high community centrality in case of *E. coli* and a continuous high community centrality plateau with two local maxima in case of *B. aphidicola* (cf. Supplementary Figures 6 and 7). These observations are in agreement with both our preliminary data derived from the visual inspection of the top 40% of reactions [Mihalik and Csermely, 2011], with other results showing that environmental variability induces a higher level of modularization in metabolic networks [Kreimer *et al.*, 2008; Parter *et al.*, 2007; Samal *et al.*, 2011] and with the recent finding that the fraction of active reactions is smaller in metabolic networks evolved to optimize a specific metabolic task [Lee *et al.*, 2012]..

We used the ModuLand Cytoscape plug-in to analyze the overlapping module structure of the metabolic networks of *Escherichia coli* and *Buchnera aphidicola*. The ModuLand plug-in calculated the Spearman's rank correlation values of each module assignment vector pair, and visualized the histogram of correlation values. The highest correlation values were 0.687 and 0.468 in the metabolic networks of *E. coli* and *B. aphidicola*, respectively. Since there were no highly correlated modules, we choose not to merge any module pairs. The plug-in also generated higher hierarchical levels of the metabolic network modules by taking the modules of the lower level networks as meta-nodes of the next level networks, and the module overlaps of the lower level networks as meta-edges of the next level. On the higher levels there was only one module in case of both organisms, so in this case no further hierarchical levels were analyzed.

Modular structures of the two metabolic networks are shown on Supplementary Figures 4 and 5. *E. coli* and *B. aphidicola* metabolic networks had 23 and 8 modules, respectively (a difference of 188%). The difference in the number of modules is even more pronounced (192%), if we compare the effective number of modules (giving an approximation to the number of modules without overlaps), which was 6.2 and 2.1 in case of the *E. coli* and *B. aphidicola* metabolic networks, respectively (see Supplementary Table 7). The larger different average module sizes (12.78 *versus* 23.75 of *E. coli* compared to *B. aphidicola*) also suggests a more differentiated module structure of *E. coli* than that of *B. aphidicola* in agreement with earlier findings [Kreimer *et al.*, 2008; Lee *et al.*, 2012; Mihalik and Csermely, 2011; Parter *et al.*, 2007; Samal *et al.*, 2011].

Since the two metabolic networks significantly differed in size, we also analyzed the main modular properties of 1000 random sample networks of the *E. coli* metabolic networks having the same number of nodes or edges like the *B. aphidicola* metabolic network. Random sample networks were constructed as described in Methods. Modularization of these 1000 random sample sub-networks was performed using specific scripts running the ModuLand binary programs packaged to the ModuLand Cytoscape plug-in. This analysis shows that the differences in the module number and module size related parameters are not caused by the different size of the *E. coli* and *B. aphidicola* networks (see Supplementary Table 7).

Size distribution of the modules was quite different in case of the two organisms (see Supplementary Figures 4 through 7). In *B. aphidicola* metabolic network two largest central modules were found around ATPS4rpp (ATP synthase) and GLCptspp (D-glucose transport via PEP:Pyr PTS). Any of these 2 largest central modules contained more nodes than the union of the rest of the modules. In case of *E. coli* no such central modules were detected having more nodes than the union of the rest of the modules. ATP-synthase was located in the central region of the largest *E. coli* module, while the 3 next largest (more-less equally sized) *E. coli* modules were found around PYK (pyruvate kinase), DRPA (deoxyribose-phosphate aldolase) and ASPTA (aspartate transaminase). The centrality of pyruvate metabolism in *E. coli* is in agreement with earlier findings [Guimera and Amaral, 2005], and GLCptspp (the central node of the other large *B. aphidicola* module) is also located in the central region of the PYK module of the *E. coli* network. The high community centrality of ATP synthase in both organisms reflects its high involvement of the communication of its direct and more distant network neighbourhood. However, the module size differences *per se* may not be considered as a measure of importance/essentiality.

To check the similarity of our *E. coli* metabolic modules with those determined before by Guimera and Amaral [2005] first we converted the reactions of the current networks to metabolites. We restricted this analysis to the substrates and products of the reactions belonging to the 10 metabolites forming the module cores, since these reactions are the most characteristic to the community structure determined by the current plug-in. We compared the metabolites present in both this model and in the KEGG modules in the supplement of Guimera and Amaral [2005]. We got 162 or 139 common intra- or inter-modular metabolites, respectively, while the number of differentially assigned intra- and inter-modular metabolites was 12 and 442, respectively. (Note that the latter number is high, since we took only the module core of the current plug-in into consideration in this analysis.) Even with these dissimilar initial conditions the two modularizations had a significant ($p=1.4 \times 10^{-7}$) overlap when using the Fisher's exact test. This shows that the two, different modularization techniques identify a statistically similar community structure.

We also compared the average homogeneity of metabolic functionalities in the modules of the two networks. For each module we chose the top ten reactions having the highest module assignment value and calculated the average number of subsystems to which these reactions were assigned (subsystem annotation was as in the published metabolic network reconstruction). The average number of subsystems in case of *E. coli* modules was 0.53, while the same average for *B. aphidicola* modules was 0.67 (the difference is significant, p = 0.0392, using the bootstrap method [Efron and Tibshirani, 1994]; where the bootstrap method was used, since the number of *B. aphidicola* modules was less than 10, which precluded the use of the Brunner-Munzel test). This finding shows that metabolic modules of *B. aphidicola* corresponded to significantly more metabolic functions than *E. coli* modules. This conclusion is in agreement with earlier findings [Guimera and Amaral, 2005; Parter et al., 2007] comparing the heterogeneity of the KEGG pathway classification of structural modules in the two organisms.

To verify our conclusion, we used the same bootstrap method [Efron and Tibshirani, 1994] on the 1000 random sample sub-networks selected from the *E. coli* network in order to have networks with the same node or edge number as can be found in the *B. aphidicola* network. We had very similar average subsystem number in the *E. coli* random sample sub-networks (0.501 and 0.508 when we selected the random *E. coli* sub-networks having an equal number of nodes or edges like those of the *B. aphidicola* network, respectively) than in the original *E. coli* network (0.53). The difference between the average values in case of the *E. coli* samples and the original *E. coli* network is not significant (two-sided p-values are 0.409 and 0.533 for the node-similar and edge-similar case), while the same difference between the *E. coli* sample sub-networks and the *B. aphidicola* metabolic network remained significant (two-sided p-values were 0.021 and 0.0294).

We also created the sub-networks of the 103 common metabolites (see Supplementary Table 8) in the two organisms. These sub-networks can be taken as alternative samples of the two metabolic networks with the same number of nodes, and they are showing the same patterns as the module structures of the original networks. The common nodes in case of *B. aphidicola* built 17 components with a giant component containing 72 nodes (see Supplementary Figure 8), while the common nodes in *E. coli* network formed a more disjoint structure with 52 smaller components, where the largest component had only 18 nodes (see Supplementary Figure 9).

For a further verification we checked, if the skewed distribution of module size causing the existence of the large centre (in form of a twin of central modules) in the *B. aphidicola* network may significantly contribute to the higher average subsystem number. It is a logical assumption that the twin central modules may distort the average value, because of their size and central position. To check their effect, first we identified the *E. coli* modules corresponding to the twin central *B. aphidicola* modules (having the module centres of ATP-synthase and glucose permease, respectively). We calculated the Spearman's rank correlation values between the module assignment vectors of the common nodes for each module-pairs and choose the seven *E. coli* modules (having module centres in the *E. coli* network: PYK, DRPA, TMDPP, ACt2rpp, ASPTA, ASPt2pp, FRD3) which have higher than 0.3 correlation with any of the twin central *B. aphidicola* modules. In the next step we generated two module core lists (listing the 10 metabolites of each module having its largest module assignment values), where we excluded the twin central modules in case of *B. aphidicola* and the corresponding 7 modules in case of *E. coli*. The difference remained significant between the residual modules, too: 0.48 for *E. coli* and 0.67 for *B. aphidicola* (using bootstrap method [Efron and Tibshirani, 1994], two-sided p-value: 0.0486). We also created a hybrid model where the twin central modules of *B. aphidicola* were 'substituted' by the corresponding 7 *E. coli* modules. The average subsystem value was 0.65 for this hybrid model, which was still significantly higher than the 0.53 we calculated for the original *E. coli* network (Brunner-Munzel test [Brunner and Munzel, 2000], one-sided p-value: 0.0029; here both systems had a larger number of modules than 10, which allowed the use of this test). These tests are suggesting that the modules of *B. aphidicola* and *E. coli* metabolic networks are organized differently in terms the homogeneity of metabolic functionalities and this difference is not due to their different size or to the existence of the twin central modules of the *B. aphidicola* network.

These results indicated that modules of the metabolic network of an organism from a variable environment (*E. coli*) are more specialized than metabolic network modules of a symbiont having a constant environment (*B. aphidicola*). It is noteworthy that our result is in agreement with earlier findings using non-overlapping modularization [Parter et al., 2007], which is a further indication after our results on protein structure network and former studies on interactomes and chromatin networks ([Mihalik and Csermely, 2011] and Sandhu, K.S., Li, G., Poh, H.M., Quek, Y.L.K., Sia, Y.Y., Peh, S.Q., Mulawadi, F.H., Sikic, M., Menghi, F., Thalamuthu, A., Sung, W.K., Ruan, X., Fullwood, M.J., Liu, E., Csermely, P. and Ruan, Y. Large scale functional organization of long-range chromatin interaction networks, submitted for publication) that the module cores reflect well the biologically relevant function of modules. As a potential mechanism of the divergence in module specialization between the two organisms, during the simplification of the *B. aphidicola* genome by the adaptation to the symbiosis [Pál et al., 2006] modules might coalesce and become more multi-functional. This hypothetical scenario, however, needs further verification.

**Comparing the Moduland plug-in to other clustering plug-ins available for Cytoscape**
Numerous methods were published for determining overlapping clusters [see e.g. Adamcsek *et al.*, 2006; Ahn *et al.*, 2010; Fortunato, 2010; Kovacs *et al.*, 2010; Palla *et al.*, 2005 and references therein]. Moreover, several very useful plug-ins are available for Cytoscape to perform discrete modularization/clustering on networks (see Supplementary Table 9). Some of them (like clusterMaker or GLay) contain multiple well-known algorithms and unify them on a single user interface providing various visualization methods. The MCODE and MINE plug-ins are based on local density measures, and therefore are useful to determine and explore clusters quickly. Some plug-ins contain overlapping modularization methods, and the two latter methods are faster than the ModuLand method in case of large networks. However, the authors are not aware of Cytoscape plug-ins, which focus on overlapping module assignment, assign each node of the network to each identified module with different intensities, determine several layers of module hierarchy and calculate various network measures based on extensively overlapping modules. In the next paragraphs we will give a short summary about all Cytoscape plug-ins for modularization/clustering we identified (see also Supplementary Table 8). More details can be found on the linked homepage of each plug-in.

*GLay plug-in, homepage*: http://brainarray.mbni.med.umich.edu/glay. GLay [Su *et al.*, 2010] offers an assorted collection of community analysis algorithms and layout functions. Some variants [Wakita and Tsurumi, 2007] of the modularity measure-based [Newman and Girvan, 2004] Clauset-Newman-Moore algorithm [Clauset *et al.*, 2004] are implemented in Java, so they can be used independently from the platform, while many other algorithms (e.g. Walk Trap [Pons and Latapy, 2006], Label Propagation [Raghavan *et al.*, 2007], Spin Glass [Reichardt and Bornholdt, 2006] and Leading Eigenvector [Newman, 2007]) are used from the igraph library[5] implemented in C language and can be run only on Windows platform.

*clusterMaker plug-in, homepage*: http://www.cgl.ucsf.edu/cytoscape/cluster/clusterMaker.html. The clusterMaker plug-in [Morris *et al.*, 2011] integrates many different clustering techniques and makes them available on a single interface. The current implementation supports clustering algorithms like k-medoid [Sheng and Liu, 2006] or hierarchical and k-means [Bishop, 1995]. The output of these methods can be displayed as hierarchical groups of nodes or as heat maps. The plug-in also supports Markov clustering [Enright *et al.*, 2002], transitivity clustering [Wittkop *et al.*, 2010], affinity propagation [Frey and Dueck, 2007], MCODE [Bader and Houge, 2003], community clustering (a CNM variant from the GLay plug-in), SCPS [Nepusz *et al.*, 2010], and also AutoSOME [Newman and Cooper, 2010] for partitioning networks based on similarity or distance values.

---
[5] igraph library: http://igraph.sourceforge.net/

*MCODE plug-in, homepage*: http://baderlab.org/Software/MCODE. The MCODE is a relatively fast clustering method [Bader and Houge, 2003], based on vertex weighting by local neighbourhood density and outward traversal from a locally dense seed node to isolate the dense regions according to given parameters. In the plug-in, the user has the possibility to fine-tune the clusters of interest (to increase or decrease the cluster size limit) without considering the rest of the network.

*MINE plug-in, homepage*: http://chianti.ucsd.edu/cyto_web/plugins/displayplugininfo.php?name=MINE. The MINE algorithm [Rhrissorrakrai and Gunsalus, 2011] was developed to discover high quality modules of gene products within highly interconnected biological networks. MINE is an agglomerative clustering algorithm very similar to MCODE, but it uses a modified vertex weighting strategy and can factor in a measure of network modularity, both of which help to define module boundaries by avoiding the inclusion of spurious neighbouring nodes within growing clusters.

*NeMo plug-in, homepage*: http://128.220.136.46/wiki/baderlab/index.php/NeMo. NeMo is a Cytoscape plug-in for unweighted network clustering. The method [Rivera *et al.*, 2010] combines a specific neighbour-sharing score with hierarchical agglomerative clustering to identify diverse network communities. NeMo is based on a score that estimates the likelihood that a pair of nodes has more common neighbours than expected by chance.

# Supplementary References